\documentclass[authoryear,final,3p,times]{elsarticle}

\usepackage{amssymb}
\usepackage{amsmath}
\usepackage{natbib}
\usepackage{appendix}
\usepackage{upgreek}
\usepackage{url}
\usepackage{lineno}
\journal{Icarus}

\begin{document}

\begin{frontmatter}



\title{Detection of propadiene (CH$_2$CCH$_2$), propene (C$_3$H$_6$) and non-detection of propane (C$_3$H$_8$) in Jupiter's northern polar stratosphere}

\author[jpl]{James A. Sinclair} 
\author[swri]{Thomas K. Greathouse}
\author[swri]{Rohini S. Giles}
\author[jpl]{Keeyoon Sung}
\author[goddard]{Conor A. Nixon}
\author[yale]{Nicholas A. Lombardo}
\author[marseille]{Vincent Hue}
\author[ssi]{Julianne I. Moses}
\author[leic]{Leigh N. Fletcher}
\author[oxford]{Patrick G. J. Irwin}
\author[jpl]{Glenn S. Orton}

\affiliation[jpl]{organization={Jet Propulsion Laboratory/California Institute of Technology},
            addressline={MS 183-301, 4800 Oak Grove Drive}, 
            city={Pasadena},
            postcode={91109}, 
            state={CA},
            country={United States}}

\affiliation[swri]{organization={Southwest Research Institute},
            addressline={M6220 Culebra Road}, 
            city={San Antonio},
            postcode={78238}, 
            state={TX},
            country={United States}}

\affiliation[goddard]{organization={Planetary Systems Laboratory, NASA Goddard Space Flight Center},
            addressline={8800 Greenbelt Road}, 
            city={Greenbelt},
            postcode={20771}, 
            state={MD},
            country={United States}}

\affiliation[yale]{organization={Department of Earth and Planetary Sciences, Yale University},
            addressline={210 Whitney Avenue}, 
            city={New Haven},
            postcode={06511}, 
            state={CT},
            country={United States}}
            
\affiliation[marseille]{organization={Aix-Marseille Université, CNES, Institut Origines, LAM},
            city={Marseille},
            country={France}}
            
\affiliation[ssi]{organization={Space Science Institute},
            addressline={4765 Walnut Street, Suite B}, 
            city={Boulder},
            postcode={80301}, 
            state={CO},
            country={United States}}

\affiliation[leic]{organization={School of Physics \& Astronomy, University of Leicester},
            addressline={University Road}, 
            city={Leicester},
            postcode={LE1 7RH}, 
            country={United Kingdom}}
            
\affiliation[oxford]{organization={Department of Physics, University of Oxford},
            addressline={Parks Road}, 
            city={Oxford},
            postcode={OX1 3PU}, 
            country={United Kingdom}}

\newpage
\begin{abstract}

\vspace{-25pt}
We report the first detection of stratospheric propadiene (CH$_2$CCH$_2$) and propene (C$_3$H$_6$) at Jupiter’s mid-to-high northern latitudes using IRTF-TEXES measurements recorded on March 5-6, 2025.  Using radiative transfer software to quantitatively test for the presence of propadiene and propene, we report a $>$12-$\sigma$ detection of propadiene and a $>$17-$\sigma$ detection of propene at high latitudes inside Jupiter’s auroral region, where the species are most concentrated. For example, at 62$^\circ$N (planetocentric) inside Jupiter’s northern auroral region (henceforth ‘NAR’), we derive a 1-mbar propadiene abundance of 2.0 $\pm$ 0.2 ppbv, which is 40 $\pm$ 3 higher than abundances predicted by the \citet{moses_2017} photochemical model (henceforth ‘MP17’), and significantly higher than the 1.2-pbbv upper limit abundance derived at 42$^\circ$N (the lowest latitude sampled by the observations). Similarly, we derive a 1-mbar propene abundance 8.1 $\pm$ 0.5 ppbv at 62$^\circ$N inside Jupiter’s NAR, which is 28 $\pm$ 2 higher than the MP17 predicted abundance and significantly higher than the 6-ppbv 1-mbar upper limit abundance derived at 42$^\circ$N.  The fact that propadiene and propene are most enriched inside Jupiter’s NAR strongly suggests that perturbations to the chemistry by auroral-related heating and exogenous ions/electrons are responsible for their significant enrichment, as has been observed for other unsaturated/aromatic hydrocarbon species. Spectral features of propane were not detected at any of the locations sampled by the data (poleward of 42$^\circ$N): 3-$\sigma$ upper limits of $\sim$10 ppbv at 10 mbar were derived at 62$^\circ$N inside Jupiter’s NAR, which is $\sim$2.5 times the MP17 predicted abundance. The non-detection of propane could, in part, be explained by the vertical sensitivity of its mid-infrared emission lines to deeper pressures, where there is negligible auroral-related heating to warm the line forming region. The results of this work strongly advocate for development of ion-neutral chemistry models of Jupiter’s polar stratosphere to quantify how strong auroral-related heating and magnetospheric particles modify the reaction pathways that produce higher-order hydrocarbons.

\end{abstract}

\begin{highlights}
\item Propadiene and propene detected for the first time at Jupiter’s mid-to-high latitudes.
\item Abundances of propadiene and propene peak inside Jupiter’s northern auroral region. 
\item Abundances of propadiene and propene are significantly enriched compared to existing models. 
\item Propane was not detected and stringent upper limit abundances were derived. 
\item Auroral-related chemistry preferentially enriches complex, unsaturated hydrocarbons.

\end{highlights}


\end{frontmatter}


\section{Introduction}

The atmospheres of the giant planets, and Saturn's moon, Titan, are host to a rich photochemistry.  CH$_4$ is transported from the deeper atmosphere, photolyzed by solar ultraviolet radiation at higher altitudes and then recombines in different ways to produce higher-order hydrocarbons \citep{gladstone_1996,moses_2005,moses_2017,hue_2018}.  These include acetylene (C$_2$H$_2$) and ethane (C$_2$H$_6$), which have strong and readily detectable spectral features on all giant planets (e.g. \citealt{kim_1985,orton_1987,nixon_2007,howett_2007,greathouse_2011,sinclair_2013,melin_2018,roman_2020}). 

Near the magnetic poles of each planet, the stratospheric photochemistry is made more complex by the influence of auroral processes.  This is particularly extreme on Jupiter since it has the strongest planetary magnetic field in our Solar System and a volcanically-active moon that serves as as an internal plasma source (e.g. \citealt{bagenal_2017, bonfond_2012}).  Magnetospheric dynamics and interactions with the solar wind ultimately drive ions and electrons deep into its neutral atmosphere producing auroral emissions from X-ray to radio wavelengths (e.g. \citealt{dunn_2017,grodent_2018,collet_2025}).  A significant amount of auroral energy reaches as deep as Jupiter's lower stratosphere ($\sim$10 mbar), which heats the atmosphere by up to several tens of Kelvin (e.g \citealt{sinclair_2017a,sinclair_2018a,sinclair_2023b,ovalle_2024}).  The auroral-related heating alone modifies chemical reaction rates thereby modulating the chemistry compared to elsewhere on the planet.  The influx of ions and electrons from the magnetosphere significantly increase the rates of ion-neutral and electron recombination reactions, which preferentially enhances the production of heavier, complex hydrocarbons (e.g. \citealt{wong_2000,friedson_2002,wong_2003,Hue2024}).  We henceforth refer to the combined effects of auroral-related heating and the influx of ions and electrons on the polar chemistry as `auroral-related chemistry'. Detailed discussion of Jupiter's polar heating and chemistry is provided in \citet{Hue2024}. 

The auroral-related heating together with the enriched abundances of hydrocarbon species make their mid-infrared emission features highly observable.  \citet{caldwell_1980} first detected the enhanced emissions of stratospheric CH$_4$ at Jupiter's poles, which were interpreted to result from auroral-related heating.  Enhanced emissions of C$_2$H$_2$, C$_2$H$_6$ and CH$_3$D (deuterated methane) were also observed in Voyager-IRIS (Infrared Interferometer Spectrometer and Radiometer, \citealt{hanel_1980}) spectra of Jupiter's poles recorded during the Voyager flybys in 1979 \citep{kim_1985}.  These same Voyager-IRIS observations of Jupiter's northern auroral region also allowed the first detection of ethylene (C$_2$H$_4$), methylacetylene (CH$_3$C$_2$H), and benzene (C$_6$H$_6$).  These species were later detected outside Jupiter's auroral region \citep{fouchet_2000,bezard_2001_c2h4,bezard_2001_c6h6} using ISO-SWS (Infrared Space Observatory's Short Wave Spectrometer, \citealt{kessler_1996}).  The enhanced mid-infrared emissions of the aforementioned species in Jupiter's auroral regions continued to be reported using ground-based instrumentation (e.g. \citealt{drossart_1986}, \citealt{livengood_1993}, \citealt{kostiuk_1993}, \citealt{romani_2008}) and from Cassini-CIRS (Composite Infrared Spectrometer, \citealt{kunde_1996}) observations recorded during the 2000-2001 flyby of Jupiter \citep{flasar_2004_jet,kunde_2004}. However, given that the strength of the emission features depend both on the temperature of the line-forming region as well as the abundance of the emitting molecule, it remained a challenge to interpret whether the enhanced hydrocarbon emissions from Jupiter's auroral regions were the result of heating, an enhanced abundance, or some combination of both.  

\citet{sinclair_2017a} performed a retrospective radiative transfer analysis of Voyager-IRIS and Cassini-CIRS data in an attempt to disentangle the contribution of heating and chemistry to the observed spectral features of Jupiter's auroral regions. Under the assumption that the vertical profile of CH$_4$ is horizontally homogenous, the emission features of stratospheric CH$_4$ were inverted using radiative transfer software to retrieve the vertical temperature profile and then, the emission features of the higher-order hydrocarbons were inverted to retrieve their vertical abundance profiles.  This analysis demonstrated that Jupiter's polar stratosphere was subject to strong auroral-related heating and C$_2$H$_2$ abundances approximately twice as high compared to non-auroral regions.  However, abundances of C$_2$H$_6$ were tentatively lower inside Jupiter's auroral region compared to elsewhere on the planet.  A similar method was adopted in further reanalyses of Voyager-IRIS, Cassini-CIRS and IRTF-TEXES (Texas Echelon Cross Echelle Spectrograph on NASA's Infrared Telescope Facility, \citealt{lacy_2002}) observations recorded in 2014 to demonstrate that Jupiter's auroral regions were also respectively enriched in C$_2$H$_4$, CH$_3$C$_2$H and C$_6$H$_6$ by factors of $\sim$1.6, $\sim$3.4 and $\sim$16 compared to photochemical models \citep{sinclair_2018a,sinclair_2019b}.  However, the aforementioned assumption that the vertical profile of CH$_4$ is horizontally homogenuous was subsequently demonstrated to be poor.  The CH$_4$ homopause, which marks the level in the atmosphere where the eddy and molecular diffusion coefficients are equal and is a commonly-used metric to quantify the rate of vertical transport, was demonstrated to be lower in pressure or higher in altitude in Jupiter's auroral regions compared to elsewhere on the planet (e.g. \citealt{clark_2018,sinclair_2020b,ovalle_2024,sinclair_2025}).   Thus, the relative abundances of the aforementioned hydrocarbons between auroral and non-auroral regions may differ if spatial variability of the CH$_4$ homopause had been considered.

\citet{giles_2023} demonstrated that C$_2$H$_2$ was enhanced by a factor of $\sim$3 in Jupiter's southern auroral region compared to elsewhere on the planet using Juno-UVS (Ultraviolet Spectrograph, \citealt{gladstone_uvs_2017}).  Their analysis did not rely on a temperature profile derived from an assumed vertical profile of CH$_4$.   Using JWST-MIRI (Mid-Infrared Instrument on the James Webb Space Telescope, \citealt{rieke_2015}) observations, and accounting for spatial variability of the CH$_4$ homopause, \citet{ovalle_2024} also determined that Jupiter's southern auroral region was enriched in C$_2$H$_2$ by a factor of $\sim$3 compared to non-auroral longitudes.  Further analysis of the same observations also found that: 1) C$_2$H$_6$ exhibited negligible change in abundance between non-auroral and auroral longitudes in the same latitude circle, 2) C$_6$H$_6$ was enriched by an order of magnitude in Jupiter's southern auroral region compared to elsewhere on the planet \citep{ovalle_c6h6_2024}.  

Overall, auroral-related chemistry yields differing effects on the higher-order hydrocarbons.  Presumably, this range of effects results from differences in molecular weight, number of carbon-to-carbon double/triple bonds, formation and destruction mechanisms, vertical profiles, temperature dependence of the reaction rates, sensitivity to ion chemistry, and altitude-dependent photochemical lifetimes.  The effects of the influx of ions and electrons on Jupiter's polar chemistry were modeled in \citet{wong_2000,friedson_2002,wong_2003}, where they found the ion chemistry in Jupiter's auroral regions enriched ethylene, benzene and polycyclic araomatic hydrocarbons (PAHs) but with insignificant changes to the abundances of acetylene and ethane.  However, there have been significant advancements in chemical kinetic rate information since these studies were conducted (e.g. see discussion in \citealt{moses_2017,moses_2018}).  State-of-the-art ion chemistry models of Jupiter's polar atmosphere are in development (at the time of writing).  In order to validate the output of such models and to gain a more holistic understanding of how auroral processes modify the stratospheric chemistry, we seek to detect and measure new hydrocarbon species in Jupiter's atmosphere and their contrast in abundance inside and outside Jupiter's auroral region.  

In this work, our goal was to detect the spectral features of propadiene (CH$_2$CCH$_2$, also called allene, an isomer of methylacetylene), propene (C$_3$H$_6$, also called propylene) and propane (C$_3$H$_8$). Spectral features of all three hydrocarbons have been detected on Saturn's moon, Titan \citep{maguire_1981,nixon_2013,lombardo_c3_2019,sylvestre_2015,lombardo_propadiene_2019}, and C$_3$H$_8$ has also been detected in Saturn's atmosphere \citep{greathouse_2006,guerlet_2009,fletcher_poles_2018,fletcher_saturn_2023} and so it seemed conceivable that they would be detectable on Jupiter.  These neutral C$_3$ species represent a range in photochemical stability and saturation.  C$_3$H$_8$ is fully saturated with single carbon-to-carbon bonds but forms from neutral photochemistry at lower altitudes and is relatively non-reactive and stable against photolysis.  Propadiene has two carbon-to-carbon double bonds, is produced at both low and high stratospheric altitudes, but is not the most stable isomer of C$_3$H$_4$ in Jupiter's stratosphere.  Propene has one single and one double carbon-to-carbon, is also produced at both low and high stratospheric altitudes, but has a relatively short photochemical lifetime.  These considerations are based on neutral photochemistry (e.g. \citealt{moses_2005}, \citealt{moses_2017}, \citealt{hue_2018}), whereas ion-induced auroral chemistry has not been as extensively explored to date.

\section{Observations}

We chose to focus our search for propene, propadiene and propane on Jupiter to its mid-to-high northern latitudes for several reasons.  First, at the higher zenith emission angles of Jupiter's mid-to-high latitudes as viewed from Earth, the tropospheric continuum is limb darkened and stratospheric emission lines are generally limb brightened, which increases the line-to-continuum ratio and makes detection of an emission line more likely.  Secondly, Jupiter's mid-to-high northern latitudes are host to the northern auroral oval, which extends poleward from 55$^\circ$N and covers longitudes from $\sim$140 to $\sim$240$^\circ$W (see Figure \ref{fig:coadd}).  As noted in the Introduction, the auroral-related chemistry has been observed to enrich the abundances of higher-order unsaturated/aromatic hydrocarbons including C$_2$H$_4$ (ethylene), CH$_3$C$_2$H (methylacetylene) and benzene (C$_6$H$_6$) (e.g. \citealt{kim_1985,kostiuk_1993,sinclair_2019b,sinclair_2023b,ovalle_2024,ovalle_c6h6_2024}).  This, together with the strong auroral-related heating the warms the line forming region, make this region an ideal place to search for trace, stratospheric species.  While there is an auroral hotspot in Jupiter's southern hemisphere, it does not extend as equatorward ($<$68$^\circ$S), and therefore is more challenging to view from Earth, compared to its northern counterpart. 

\subsection{Measurement strategy}

High spectral resolution (45000 $<$ R $<$ 80000) TEXES \citep{lacy_2002} spectral scans were performed on NASA's Infrared Telescope Facility on March 5-6 2025.  On March 5th 2025, spectra were recorded in settings centered at 587, 748, and 1248 cm$^{-1}$.  The 587 cm$^{-1}$ and 1248 cm$^{-1}$ spectral settings respectively capture the (readily-detectable) spectral features of the H$_2$ S(1) quadrupole and CH$_4$ emission, which are commonly inverted to derive the vertical temperature profile (e.g. \citealt{fletcher_2016,sinclair_2018a}). The 748 cm$^{-1}$ spectral setting captures the stratospheric emission lines of C$_2$H$_2$ and the targeted emission lines of C$_3$H$_8$.  On the first night of observations, focus was given to the 748 cm$^{-1}$ setting (and the 587/1248 cm$^{-1}$ settings needed for contemporaneous temperature information) to increase the effective exposure time to compensate for the poorer telluric transmission and therefore poorer signal-to-noise ratio (SNR) in this setting.  On March 6th 2025, spectra were recorded in settings centered at 587, 843, 912, and 1248 cm$^{-1}$.  The 843 cm$^{-1}$ captures predominantly the emission lines of C$_2$H$_6$, and the targeted spectral features of propadiene.  Spectra recorded at 912 cm$^{-1}$ capture the stratospheric emission features of C$_2$H$_4$, tropospheric NH$_3$ and PH$_3$, and the targeted spectral features of propene.   

\noindent In a given spectral setting, the slit of the instrument (ranging from 9.9-13.3'' in length and 1.3-1.9'' in width depending on the spectral setting) was aligned parallel to Jupiter's central meridian and scanned west to east across Jupiter's mid-to-high northern latitudes.  Scans were performed in steps of half the slit width for Nyquist sampling and with a 4-second exposure recorded at each step and chopping between the target and a blackbody card mounted in the instrument foreoptics.  Several planet-free steps on the sky were included at the beginning and end of each scan, which allowed for efficient sky subtraction.   The scans were repeated in all the remaining spectral settings for a given night.  Tables A.1 and A.2 provide details of all individual scans recorded on March 5 and 6, 2025, respectively.  Raw and calibrated forms of the data are publicly available - see Data Availability section for details. 

\subsection{Reduction \& Calibration}

For each individual scan, the sky emission was calculated from the sky steps at the beginning and end of the scan and subtracted from the target spectra.  Flatfielding was achieved using the normalized blackbody exposures.  A first-order telluric correction was performed by dividing the target spectra by (R$_{b}$ - R$_{sky}$), where R$_{b}$ and R$_{sky}$ are the radiance spectra of the foreoptics blackbody and the sky, respectively.  The absolute calibration scale factor was calculated using the blackbody exposure and its known temperature at the time of the scan and applied to all spectra of Jupiter.  Individual spectra were corrected for Jupiter's Doppler shift, which results both from the radial velocity of Jupiter with respect to Earth and the rotational velocity of Jupiter.  For each individual scan, the wavenumber-dependent noise-equivalent spectral radiance was calculated by determining the standard deviation in radiance of background sky pixels.  

\subsection{Spatial coaddition}
\begin{figure}[t!]
\begin{center}
\includegraphics[width=0.49\textwidth]{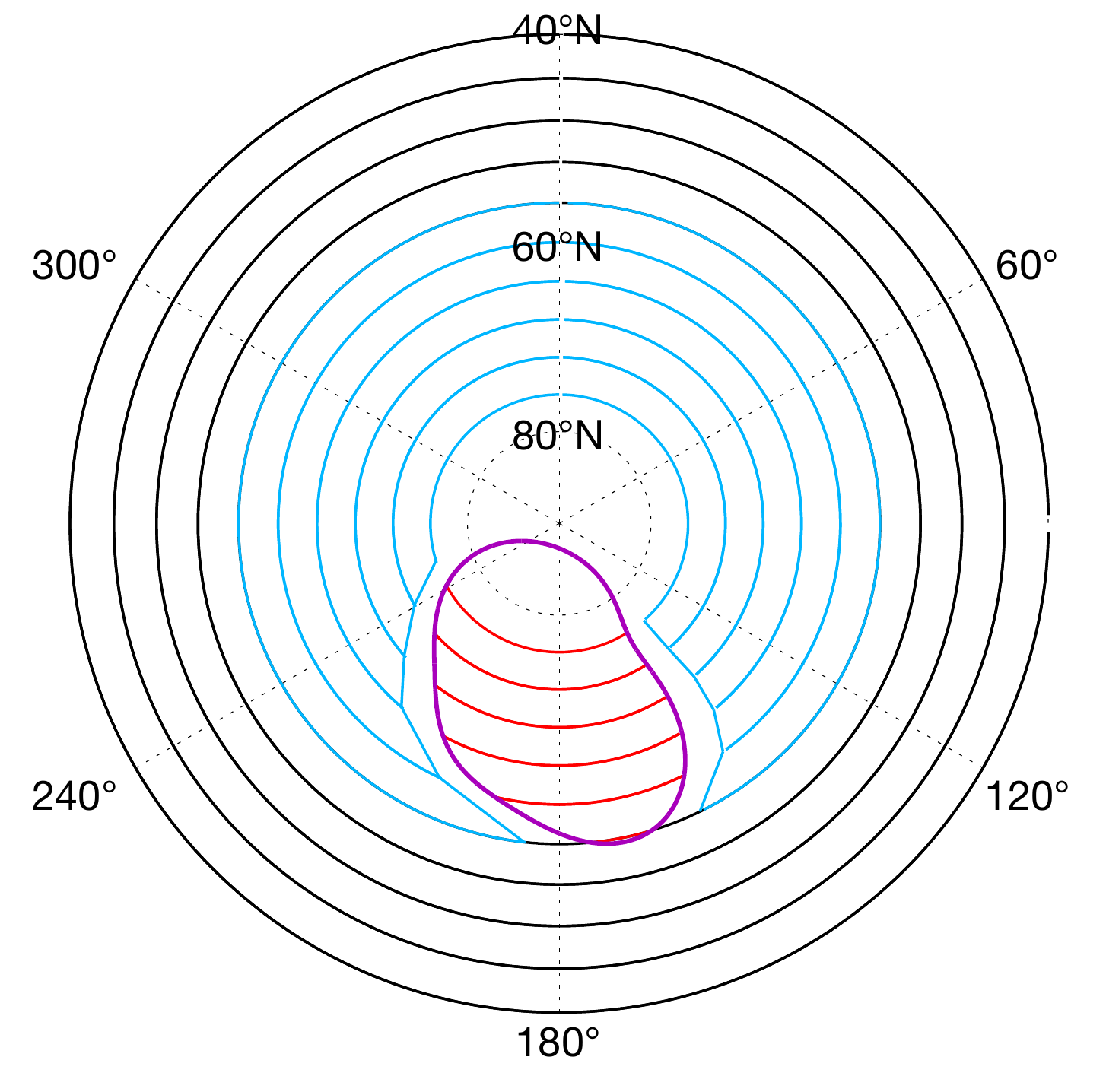}
\caption{A northern polar projection of Jupiter indicating the latitude and longitude ranges adopted for coaddition of individual spectra.  The circumference latitude is 40$^\circ$ (planetocentric), enclosed concentric circles are spaced by 4$^\circ$ in latitude.  The magenta line represents the statistical mean position of the auroral oval \citep{bonfond_2017}.  Spatial bins in black represent regions that do not overlap in latitude with the auroral oval, blue and red spatial bins represents areas outside and inside the auroral oval, respectively. TEXES achieves diffraction-limited spatial resolutions of 0.7'' to 1.4'' over the wavelength range adopted in this study, which corresponds to a latitude-longitude footprint of 5-10$^\circ$ at 60$^\circ$N.}
\label{fig:coadd}
\end{center}
\end{figure}
In order to increase the effective signal-to-noise ratio, individual spectra were spatially coadded according to the binning scheme shown in Figure \ref{fig:coadd}.  Spectra were sorted into latitude bins 8$^\circ$ in width and Nyquist-sampled by 4$^\circ$.  For latitude bins equatorward of the northern auroral oval ($<$55$^\circ$N), spectra recorded over all sampled longitudes were coadded to compute a longitudinal-mean spectrum.  For latitude bins that overlap with the auroral oval ($>$55$^\circ$N), spectra were sorted into those that sampled outside and inside the auroral oval (blue and red ranges in Figure \ref{fig:coadd}, respectively) and coadded to compute a `non-auroral mean' and `auroral mean' spectrum, respectively. We chose this binning scheme to: 1) reflect the 0.7-1.4'' diffraction-limited spatial resolutions achieved by TEXES over the wavelengths adopted in this study, which corresponds to a latitude-longitude footprint of 5-10$^\circ$ at 60$^\circ$, 2) maximize the number of spectra available for coaddition in the lower-latitude/non-auroral regions to compensate for the poorer signal-to-noise ratio in such cooler regions, 3) resolve latitudinal information both inside and outside the auroral oval, 4) ensure that spectra of very similar emission angles/geometries were coadded together.

\begin{figure}[t!]
\begin{center}
\includegraphics[width=0.8\textwidth]{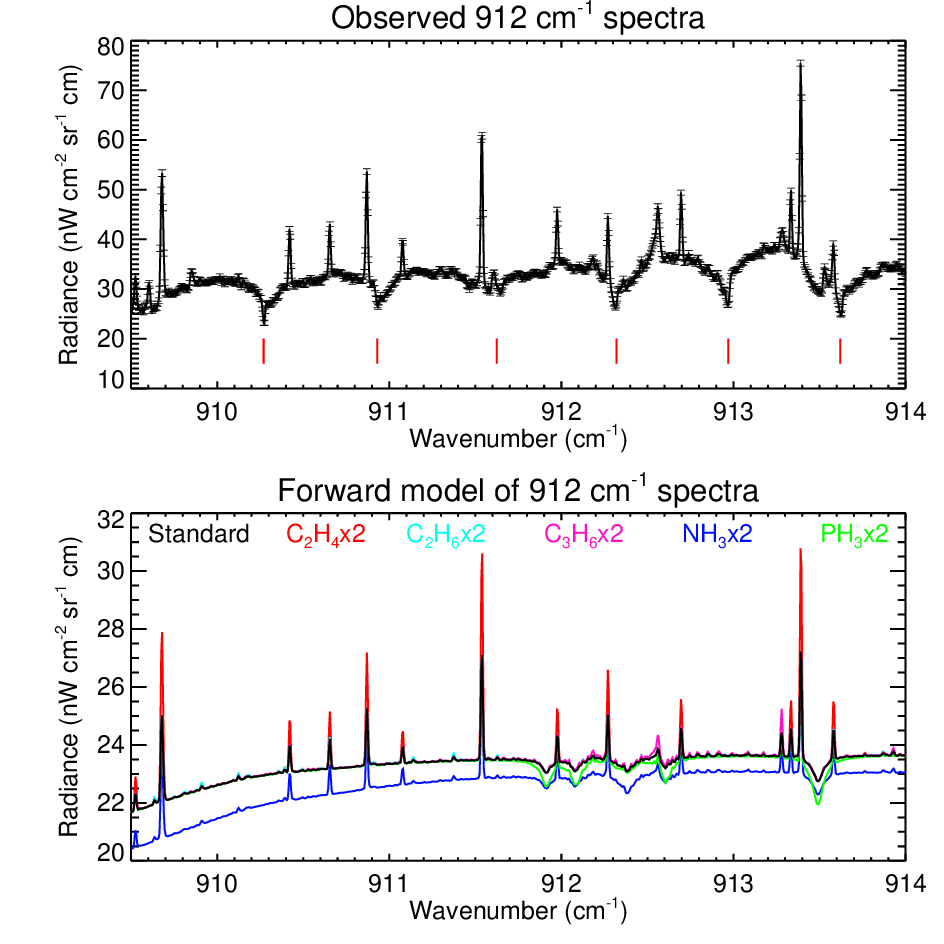}
\caption{The top panel shows an auroral mean, 912 cm$^{-1}$ spectrum (black line with error bars) at 62$^\circ$N recorded on March 6 2025.  Detector artefacts are readily identifiable as dips in radiance that are regularly spaced in wavenumber, as indicated by the vertical, red lines. The bottom panel shows synthetic spectra of Jupiter for comparison with the observed spectra.  The black spectrum was calculated assuming the nominal atmospheric model and radiative transfer model detailed in Section \ref{sec:rad_tran}.  Additional spectra are shown where a given species' abundance was increased by a factor of 2, as indicated in the legend.}
\label{fig:example_912}
\end{center}
\end{figure} 

\subsection{Noise propagation}

As noted previously, the wavenumber-dependent noise-equivalent radiance for each spectrum was calculated by determining the standard deviation in radiance of background sky pixels.  The sky emission is much more variable at wavelengths of poorer telluric transmission and thus, the noise on the target spectra is higher at such wavelengths.  This means that regions of poorer telluric transmission/higher noise are weighted less upon analysis with optimal estimation retrieval techniques (see Section \ref{sec:rad_tran}).

\begin{figure*}[t!]
\begin{center}
\includegraphics[width=0.8\textwidth]{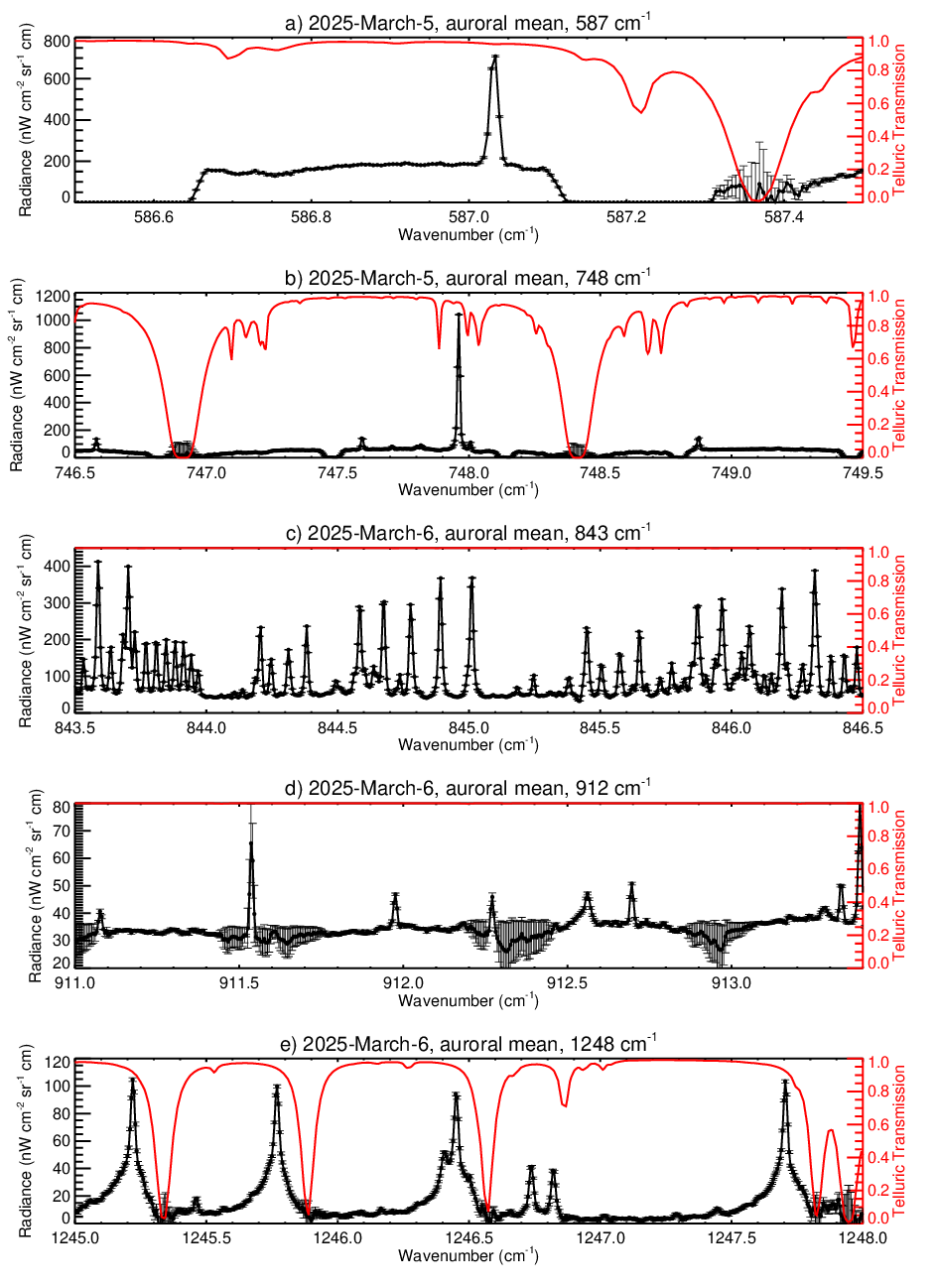}
\caption{Coadded spectra recorded on March 5th 2025 at 587 cm$^{-1}$ (1st panel), 748 cm$^{-1}$ (2nd panel), and on March 6th 2025 at 843 cm$^{-1}$ (3rd panel), 912 cm$^{-1}$ (4th panel) and 1248 cm$^{-1}$ as black lines with error bars and with radiances according to the left-hand y-axis. Telluric transmission spectra are shown as solid red lines according to the right-hand axis.  Jupiter spectra have been Doppler-shifted to the rest frame, telluric transmission spectra have been Doppler-shifted by the same amount as the Jupiter spectra.  }
\label{fig:example_texes}
\end{center}
\end{figure*} 

The noise on the coadded spectrum was assumed to be the larger of either the standard deviation on the mean or the uncertainty on the mean (i.e. $\bar{\sigma(\nu)} = \sqrt{\sigma_1(\nu)^2 + \sigma_2(\nu)^2 + ... + \sigma(\nu)_N^2}/N$, where $\sigma_1(\nu)$, $\sigma_2(\nu)$ and $\sigma_N(\nu)$ are the noise-equivalent radiance spectra for the 1st, 2nd and Nth spectra coadded).

For spectra recorded in the 912 cm$^{-1}$ setting, detector artefacts were apparent, as shown in Figure \ref{fig:example_912}.  These artefacts originate as narrow gaps in spectral coverage on the detector in individual spectra but become wider features when many spectra recorded over a range of Doppler shifts (due to Jupiter's rotation) are corrected into the rest frame and then coadded.  While standard practice is to remove the artefacts by fitting a high order polynomial, this would have been complicated by the fact that some of the artefacts were convolved with Jovian spectral features (e.g. a C$_2$H$_4$ emission line at 913.6 cm$^{-1}$ or a NH$_3$ absorption feature at 912.3 cm$^{-1}$, Figure \ref{fig:example_912}).  Instead, the noise within $\Delta\nu \sim$0.15 cm$^{-1}$ of the artefacts was increased such that they were insignificant with respect to uncertainty and therefore ignored in subsequent spectral fitting.  Figure \ref{fig:example_texes}d shows an example of a 912 cm$^{-1}$ recorded on March 6 2025 with the noise adjusted as detailed above.

\subsection{Radiometric calibration adjustment}

As noted in previous work (e.g. \citealt{sinclair_2017b,sinclair_2018a,sinclair_2023b,sinclair_2025}), we found that we could not simultaneously fit the 587 cm$^{-1}$ and 1248 cm$^{-1}$ spectral settings with the same temperature profile unless radiances in the 587 cm$^{-1}$ were increased by a factor of $\sim$1.3.  In the aforementioned past work, this has been attributed to some form of beam dilution in the 587 cm$^{-1}$ setting.  All 587 cm$^{-1}$ spectra analyzed in this study were therefore scaled by a factor of 1.3 with respect to their nominal absolute calibration.   The nominal calibration of the remaining spectral settings was found to be sufficient for spectral fitting.   Figure \ref{fig:example_texes} shows the auroral mean spectra at 62$^\circ$N in all spectral settings. 

\section{Radiative Transfer Model}\label{sec:rad_tran}

Radiative transfer modeling and inversions were performed using the NEMESIS radiative transfer code \citep{irwin_2008}.  NEMESIS is a publicly available software suite - see Data Availability section for further details. 

\subsubsection{Atmospheric model}

\begin{figure}[t!]
\begin{center}
\includegraphics[width=0.7\textwidth]{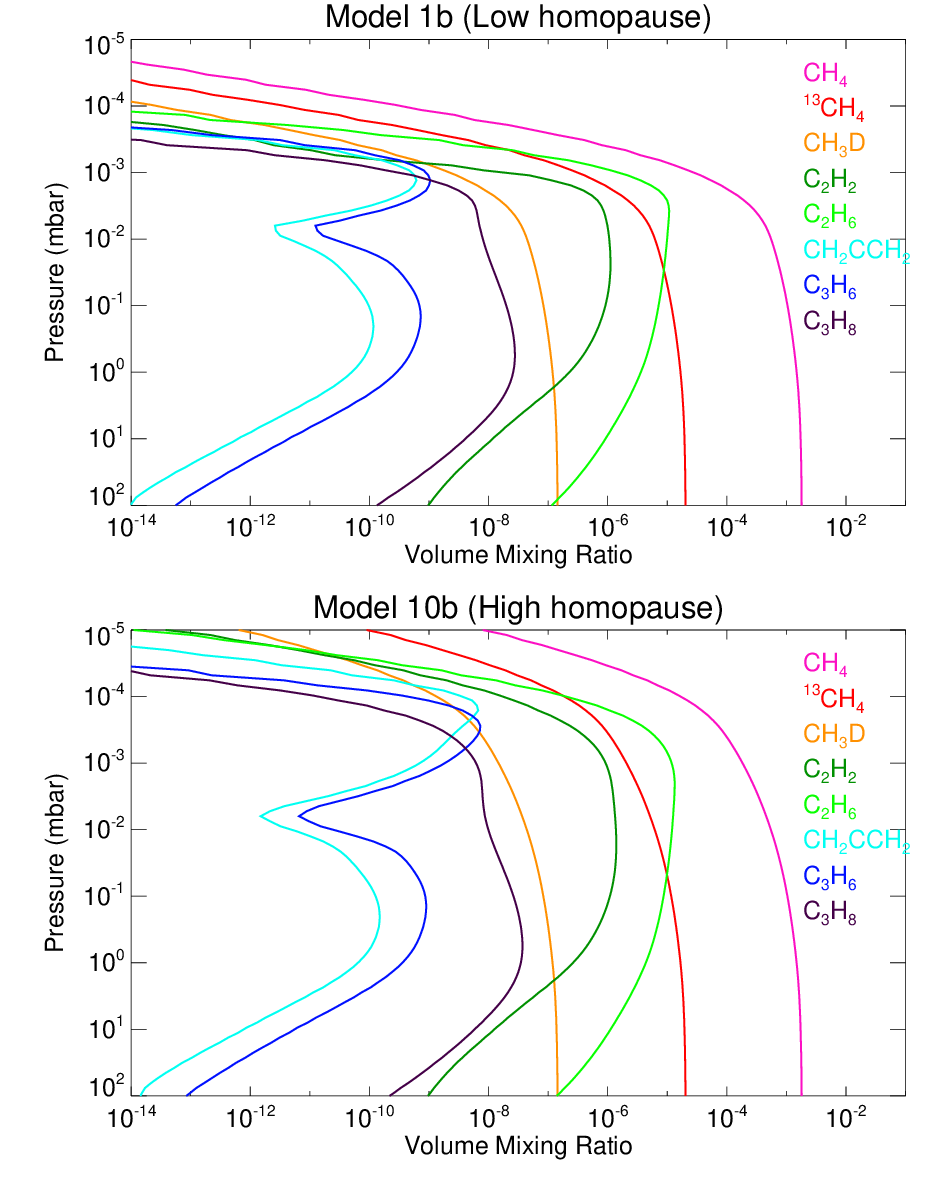}
\caption{The predicted vertical profiles of Jupiter's hydrocarbon species for a low (top panel) and high (bottom panel) homopause model.  Model 1b was adopted in fitting spectra outside of the auroral oval, Model 10b was adopted in fitting spectra inside the auroral oval.  These profiles are based on the neutral photochemical model presented in \citet{moses_2017} and are henceforth referred to as the `MP17' profiles.}
\label{fig:cxhy}
\end{center}
\end{figure}

The vertical profiles of H$_2$, He, NH$_3$ and PH$_3$ were adopted from \citet{sinclair_2020b} and references therein.  A compact, grey aerosol layer at the 800-mbar level with an optical depth of 0.3 was also assumed to be consistent with results derived for Jupiter's high latitudes in \citet{fletcher_2016}. For the vertical profiles of temperature and hydrocarbons, we adopted the photochemical models presented in \citet{sinclair_2025} and \citet{ovalle_2024}.  These photochemical models are based on those published in \citet{moses_2017} but with the model output at 60$^\circ$N and calculated over a range of homopause levels.  These models are included in a Mendeley archive linked to this publication - see the Data Availability section for further details.  As demonstrated in \citet{sinclair_2020b,ovalle_2024,sinclair_2025}, the homopause level is elevated in Jupiter's auroral regions compared to elsewhere on the planet.  For radiative transfer modeling of observations inside of Jupiter's northern auroral region, we adopted the photochemical model `10b' with a 35-nbar homopause level, which was typically found to be the best-fitting homopause level in observations sampling inside Jupiter's auroral region \citep{sinclair_2025}.  For observations sampling outside of Jupiter's northern auroral region, we adopted the `1b' model assuming a 754-nbar homopause level.  Figure \ref{fig:cxhy} shows the predicted vertical profiles of the relevant hydrocarbon species for this work. We henceforth refer to these predicted profiles by \citet{moses_2017} as the `MP17' profiles.

\begin{figure}[t!]
\begin{center}
\includegraphics[width=0.7\textwidth]{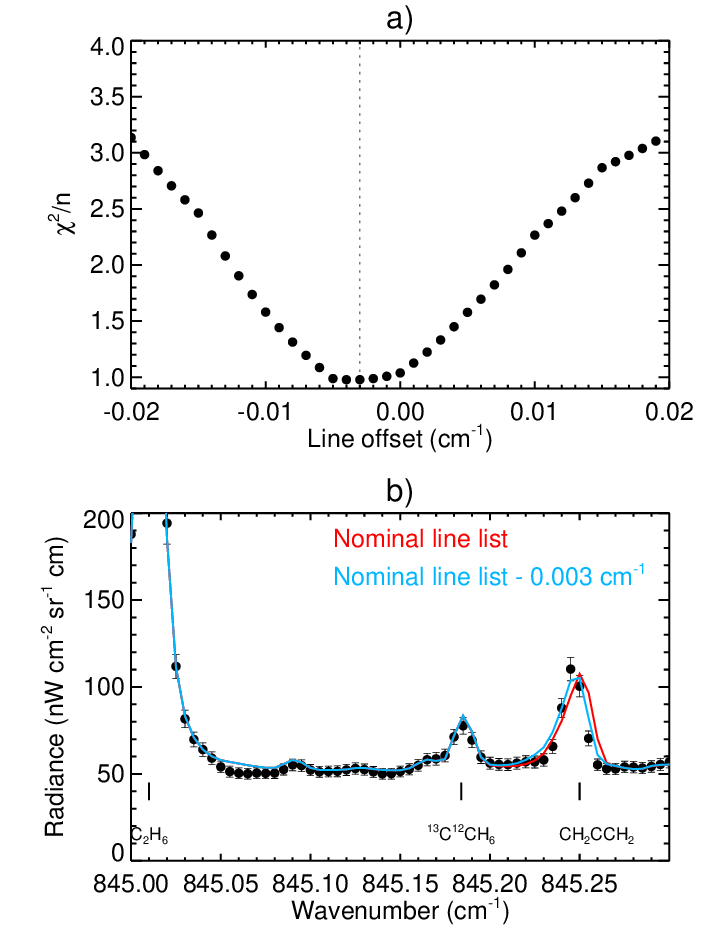}
\caption{a) The goodness-of-fit calculated over 845.2 - 845.3 cm$^{-1}$ between the observed auroral-mean spectrum at 66$^\circ$N recorded on March 6 2025 and a synthetic spectrum as a function of the wavenumber offset applied to propadiene lines between 800-900 cm$^{-1}$.  An offset of -0.003 cm$^{-1}$ produces the best correspondence between the observed and modelled location of the propadiene feature at $\sim$845.25 cm$^{-1}$, as indicated by the vertical, dashed line.  b) the observed auroral-mean spectrum at 62$^\circ$N (points with error bars), a synthetic spectrum assuming the nominal wavenumbers of propadiene lines (red) and a synthetic spectrum with an offset of -0.003 cm$^{-1}$ applied to the wavenumbers of propadiene lines.  The location of C$_2$H$_6$ lines in the spectrum are also indicated. }
\label{fig:propadiene_line}
\end{center}
\end{figure} 

\subsubsection{Spectroscopic line date}

The spectroscopic line data of H$_2$, PH$_3$, NH$_3$, CH$_4$ and its isotopologues were adopted from Supplementary Table 1 of \citet{fletcher_poles_2018}.  For C$_2$H$_2$, C$_2$H$_4$, C$_2$H$_6$, the line data was based on that of HITRAN 2020 \citep{hitran_2020} but with modifications to the line parameters for a H$_2$-dominated atmosphere.  For C$_2$H$_2$, the foreign broadened line width and temperature dependence were adjusted according to \citet{varansi_1992}.  For C$_2$H$_4$, foreign broadening line widths and temperature dependencies were modified according to the results of \citet{bouanich_2003} and \citet{bouanich_2004}.  For C$_2$H$_6$, the line broadening temperature dependence was modified to n = 0.94 for consistency with \citet{halsey_1988}.  While HITRAN 2024 \citep{hitran_2024} was released during the preparation of this article, as far as we can tell, there were no changes to the line lists of C$_2$H$_2$, C$_2$H$_4$ and C$_2$H$_6$ at the wavelengths of this study.

For propadiene, we adopted the line list from GEISA 2020 \citep{delahaye_2021} but with the temperature dependence of the air broadening halfwidth modified to 0.5 to be consistent with that assumed for methylacetylene (an isomer of propadiene) in a giant planet atmosphere (see Supplementary Table 1 of \citet{fletcher_poles_2018}).  In the absence of empirical measurements of line broadening widths for propadiene in a H$_2$-dominated atmosphere, we adopted the air broadening width.  The GEISA linelist is itself based on that compiled by \citet{lombardo_c3_2019} (see Section 2.2.1 of their paper), which includes a correction of line positions to match experimental room temperature data recorded at 0.08 cm$^{-1}$ resolutions.  In preliminary spectral fitting, we noticed that the position of the propadiene line in synthetic spectra was consistently offset from that in the observed spectra.  Figure \ref{fig:propadiene_line}a shows the goodness-of-fit between observed and synthetic spectra as a function of the offset applied to the propadiene line positions.  An offset of -0.003 cm$^{-1}$ allows the synthetic spectra to best reproduce the observed spectra, which is demonstrated spectrally in Figure \ref{fig:propadiene_line}b.  We do not attribute this offset to be a result of wavenumber calibration uncertainty in the observed spectra since the location of C$_2$H$_6$ lines in observed spectra accurately match those of synthetic spectra.  Instead, we attribute the offset to the significantly coarser resolution (0.08 cm$^{-1}$) of laboratory measurements used to update the propadiene list, as noted above.   We also note a similar offset between observed and synthetic propadiene lines in an analysis of Titan spectra \citep{lombardo_propadiene_2019}. The results presented in this paper adopted a -0.003 cm$^{-1}$ shift to all propadiene lines in the 800 - 900 cm$^{-1}$ range. 

For propene, we adopted a pseudoline list based on laboratory measurements \citep{sung_2018}.  As with propadiene, we adopted the air broadening widths in the absence of any hydrogen-based line width measurements. For propane, we similarly adopted the pseudoline list based on \citet{sung_2013} but with the foreign-broadening widths and temperature dependences respectively modified to 0.08 cm$^{-1}$ and 0.75 for all lines for consistency with the approach of \citet{fletcher_poles_2018}.

\subsubsection{Vertical sensitivity}

\begin{figure}[t!]
\begin{center}
\includegraphics[width=0.7\textwidth]{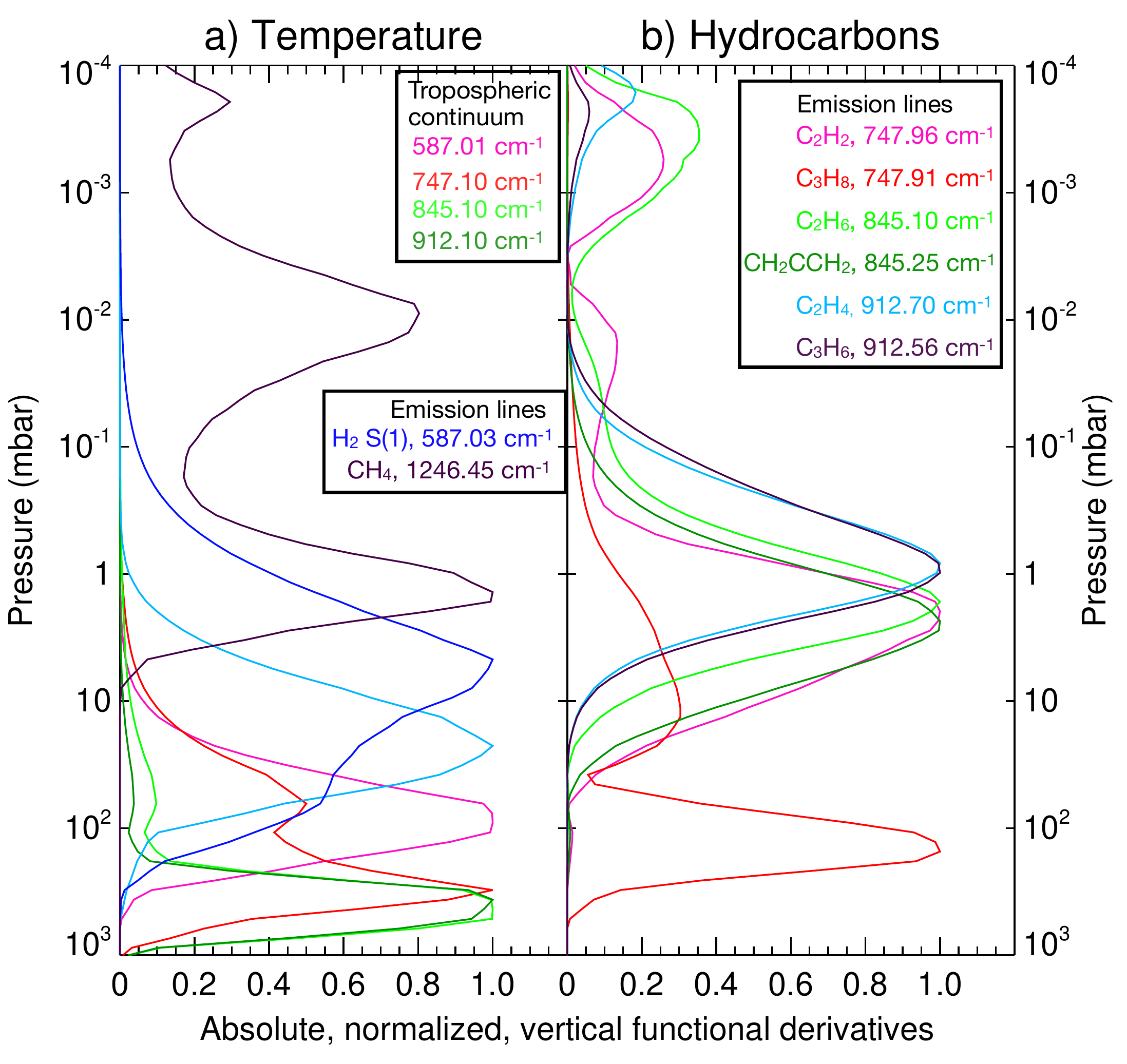}
\caption{The absolute, normalized vertical functional derivatives with respect to a) temperature and b) hydrocarbon abundances for the spectral features indicated in the legends.  }
\label{fig:contr_all}
\end{center}
\end{figure}

The spectra presented in this study capture continuum emission, the emission lines of H$_2$ S(1), CH$_4$, C$_2$H$_2$, C$_2$H$_4$, C$_2$H$_6$, and the expected location of lines of CH$_2$CCH$_2$, C$_3$H$_6$ and C$_3$H$_8$.  The vertical functional derivatives, $dR_i/dx_j$, where $R_i$ is the radiance at the i$^{th}$ wavenumber, and $x_j$ is an atmospheric parameter (e.g. temperature, abundance) at the $j^{th}$ atmospheric level, were calculated in order to determine the vertical sensitivity of the aforementioned spectral features. The calculations assumed the vertical profiles of Model 10b (Figure \ref{fig:cxhy}) and an auroral temperature profile retrieved from Gemini-TEXES observations recorded on March 19 2017 (see Figure 10 of \citet{sinclair_2023b}).    The vertical functional derivatives with respect to temperature and all aforementioned hydrocarbons are shown in Figure \ref{fig:contr_all}a and b, respectively.  

The continuum spectral regions sound Jupiter's upper troposphere from $\sim$70 to 700 mbar.  We note that the tropospheric continuum in the 748, 843 and 912 cm$^{-1}$ spectral settings sound significantly deeper pressure than the tropospheric continuum in the 587 cm$^{-1}$ setting. The emission lines of H$_2$ S(1) at 587.03 cm$^{-1}$ and CH$_4$ from 1245 - 1252 cm$^{-1}$ collectively sound the atmosphere from $\sim$80 mbar to $\sim$5 $\upmu$bar. 

The emission lines of C$_2$H$_2$, C$_2$H$_4$ and C$_2$H$_6$ peak in sensitivity in the 5 to 0.5 mbar range with a smaller, secondary peak in sensitivity at microbar pressures.  The targeted spectral features of propadiene and propene peak in sensitivity at the $\sim$3 mbar and $\sim$1 mbar levels, respectively.  The targeted spectral features of propane predominantly sound the upper troposphere at $\sim$200 mbar (which would be observed as a broad, absorption feature) with a secondary peak in sensitivity at $\sim$10 mbar (which would be observed as a narrower emission line).  

\section{Analysis}
\subsection{Retrieval of temperature distributions}\label{sec:Tp_retr}

The mid-infrared emission features of stratospheric hydrocarbons are modulated both by the temperature of the line-forming region as well as the abundance of the emitting molecule.  We therefore first derive the temperature distributions before fitting the higher-order hydrocarbon emission features to derive their abundances.

As noted previously, we respectively adopt Models 1b and 10b (Figure \ref{fig:cxhy}) in fitting spectra outside and inside Jupiter's northern auroral region.  In preliminary retrievals, we retrieved the vertical temperature profile from the H$_2$ S(1) and CH$_4$ emissions and then adopted the retrieved temperature profile in the spectral fitting of the 748, 843 and 912 cm$^{-1}$ spectral settings.  In such a two-step process, the temperature profile derived in the first step did not sufficiently fit the tropospheric continua in the 748, and 912 cm$^{-1}$ spectral settings.  This was determined to result from differing vertical sensitivity: as shown in Figure \ref{fig:contr_all}, the tropospheric continuum in the 587 cm$^{-1}$ setting peaks in sensitivity at $\sim$100 mbar whereas those of the 748, 843 and 912 cm$^{-1}$ settings are more sensitive to deeper pressures.  This issue was resolved by simultaneously fitting the spectral emission features of H$_2$ S(1) in the 587 cm$^{-1}$ setting, CH$_4$ in the 1248 cm$^{-1}$ setting and the tropospheric continua in the 748 cm$^{-1}$ or 912 cm$^{-1}$ setting.

For observations on March 5th 2025, the vertical temperature profile was retrieved by simultaneously inverting the H$_2$ S(1) quadrupole emission feature at 587.03 cm$^{-1}$, the tropospheric continuum in the 587 cm$^{-1}$ setting, the tropospheric continuum in the 748 cm$^{-1}$ spectral setting, and the CH$_4$ emissions in the 1248 cm$^{-1}$ spectral setting.  Spectra at 843 cm$^{-1}$ and 912 cm$^{-1}$ were not measured on March 5 2025.  

\begin{figure*}[t!]
\begin{center}
\includegraphics[width=0.95\textwidth]{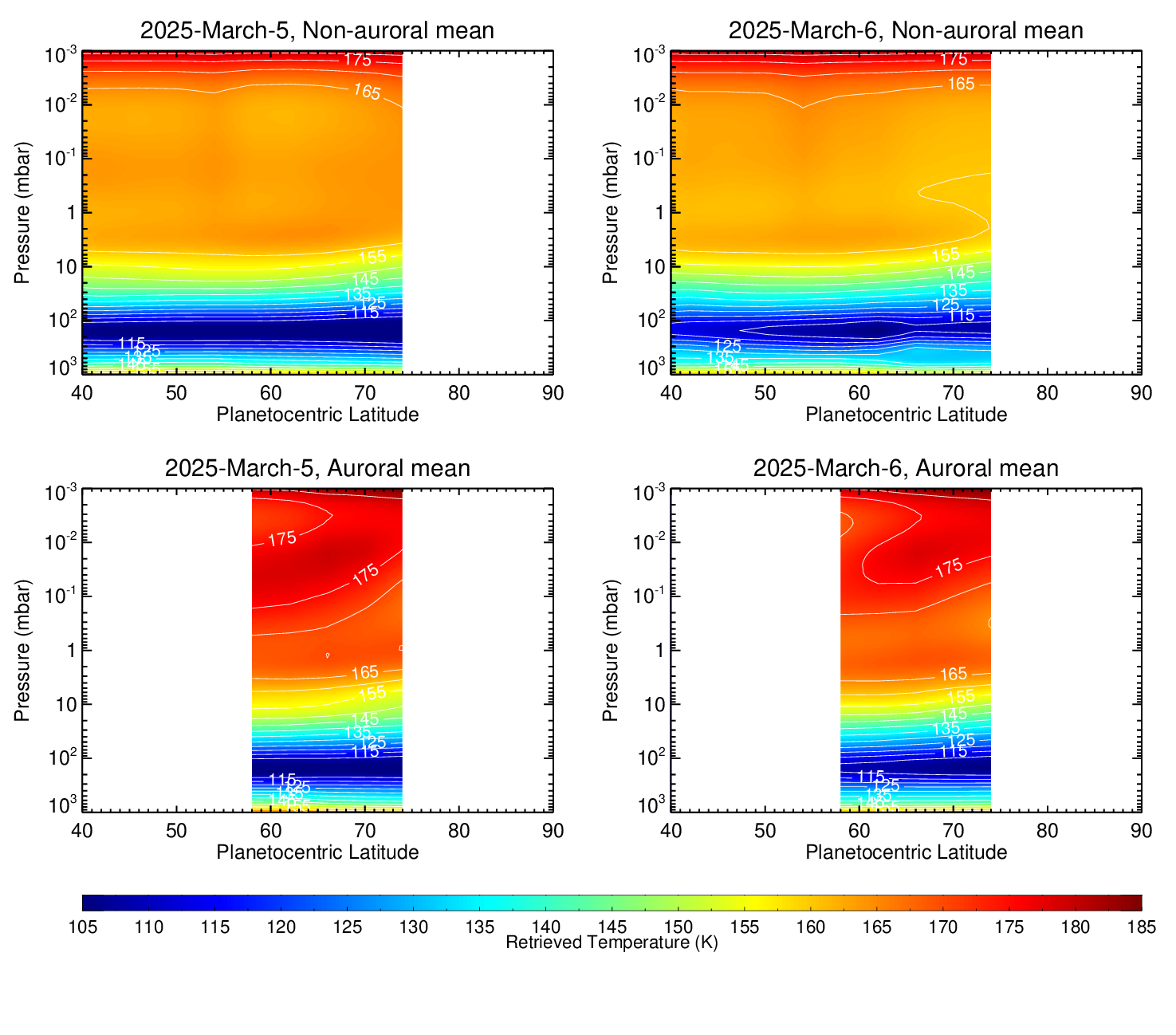}
\caption{Retrieved temperature distributions for observations recorded on March 5, 2025 (left column) and March 6, 2025 (right column).  Results for observations sampling longitudes outside and inside the northern auroral region are shown in the top and bottom rows, respectively.  Results are colored according to the colorbar at the bottom of the figure. }
\label{fig:Tp}
\end{center}
\end{figure*}

For March 6th 2025, a similar procedure was performed but using the tropospheric continua in the 843 and 912 cm$^{-1}$ spectral settings.   The tropospheric continuum in the 912 cm$^{-1}$ is also modulated by the absorption features of tropospheric ammonia (NH$_3$) and phosphine (PH$_3$) gas (see Figure \ref{fig:example_912}) and so their vertical profiles were also allowed to vary in order to optimize the fit.   The vertical profiles, $q(p)$, of gaseous NH$_3$ and PH$_3$ were parameterized by a deep, constant volume mixing ratio, $q_0$ up to a knee pressure, $p_0$ above which the abundance decayed according to a fractional scale height, $f$, as shown in the Equation below.  As in \citet{fletcher_2016}, the knee pressures of NH$_3$ and PH$_3$ were fixed to 800 mbar and 750 mbar, respectively, and only the deep volume mixing ratio and fractional scale height were varied in order to reduce the size of the parameter space. 

\[
  q(p) =
  \begin{cases}
    q_0 & p > p_0 \\
    q_0 \left( \frac{p}{p_0} \right) ^ \frac{1-f}{f} & p \le p_0 \\
  \end{cases}
\]

Figure \ref{fig:Tp} shows the retrieved temperature distributions retrieved from observations recorded on March 5 and 6 ,2025.  The presence of strong auroral heating is apparent as deep as $\sim$4 mbar in comparing retrieved temperatures inside and outside the auroral regions in a given latitude band, as noted in previous work (e.g. \citealt{sinclair_2017a,sinclair_2018a,sinclair_2023b}).   We discuss the interpretation of auroral heating further in Section \ref{sec:discuss}.  For NH$_3$, we derived deep volume mixing ratios ranging from $\sim$130 to $\sim$160 ppmv with fractional scale heights of $\sim$0.2.  This corresponds to an abundance of 20-30 ppmv at 500 mbar, which is in agreement within uncertainty with abundances derived by \citet{fletcher_2016} using Cassini-CIRS observations recorded in 2000-2001 and IRTF-TEXES observations recorded in 2014. Similarly, for PH$_3$, we derive deep volume mixing ratios and fractional scale heights of $\sim$1.3 ppmv and $\sim$0.3, respectively.  This corresponds to 500-mbar abundances of $\sim$0.5 ppmv, which is also in agreement with values derived by \citet{fletcher_2016}.  These results are not included in Figure \ref{fig:Tp} for clarity.

\subsection{Detection of propadiene (CH$_2$CCH$_2$) at 845.25 cm$^{-1}$}

The 843 cm$^{-1}$ spectra are dominated by the emission lines of C$_2$H$_6$. We therefore first inverted the 843 cm$^{-1}$ spectra by adopting the vertical temperature profile retrieved in Section \ref{sec:Tp_retr} and allowing the vertical profile of C$_2$H$_6$ to vary.  The MP17 profiles for C$_2$H$_6$ (Figure \ref{fig:cxhy}) were adopted as \textit{a priori} in inverting the 843 cm$^{-1}$ spectra.   The vertical profile of C$_2$H$_6$ was allowed to vary continuously at all altitudes but sensitivity is greatest in the 5- to 0.5-mbar pressure range (Figure \ref{fig:contr_all}).

Once the vertical profiles of temperature and C$_2$H$_6$ had been retrieved at each location, forward model spectra were computed over a grid where the MP17 profile for propadiene (Figure \ref{fig:cxhy}, \citealt{moses_2017}) was scaled by a constant at all altitudes over a range from 0 to 60.  For each forward model, the $\chi^2$ statistic (Equation \ref{eq:chi2}) was calculated from 845.22 - 845.29 cm$^{-1}$, which covers the expected position of the strongest propadiene line and surrounding tropospheric continuum.
\begin{figure*}[t!]
\begin{center}
\includegraphics[width=0.99\textwidth]{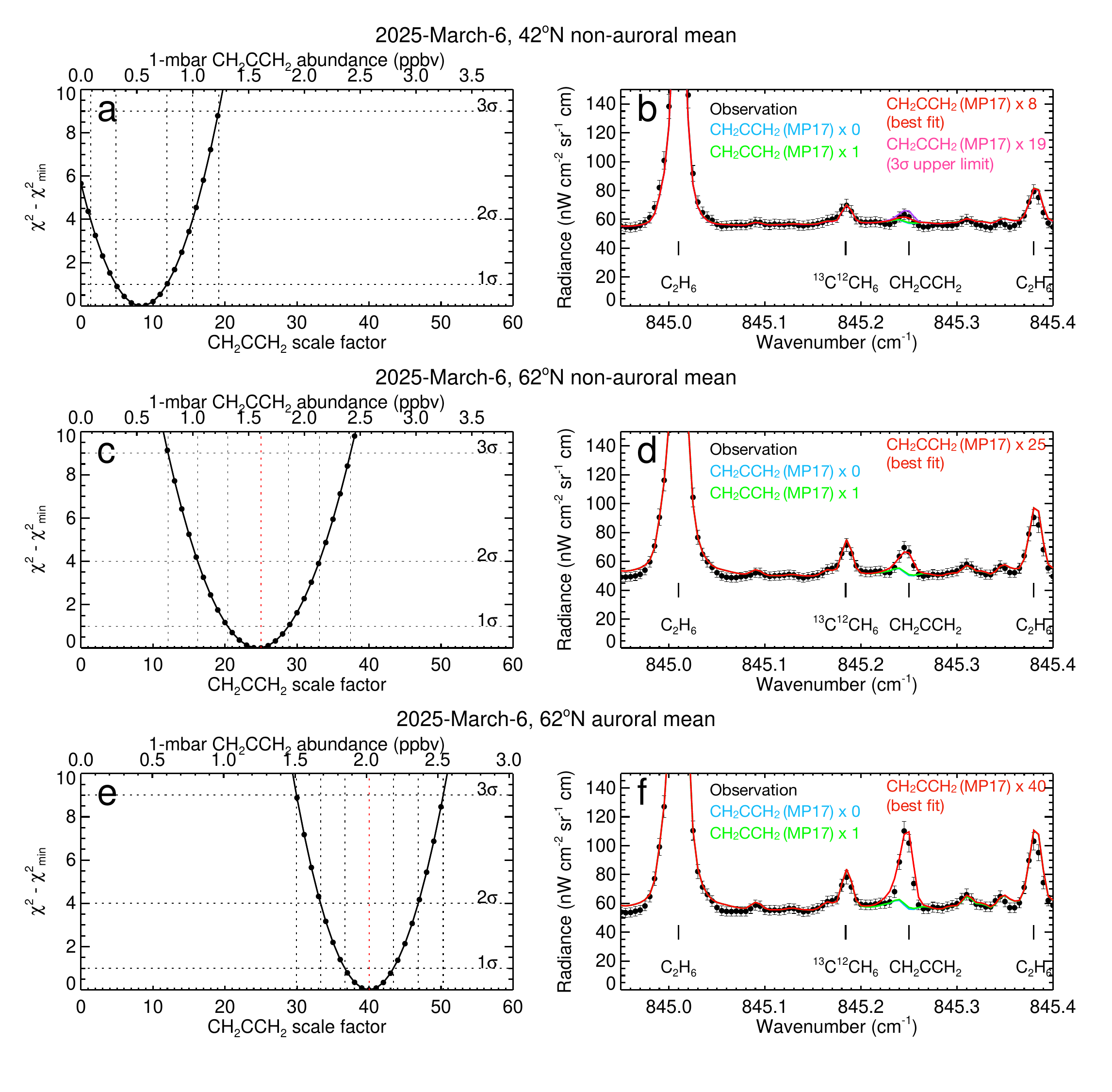}
\caption{a) shows variations in absolute $\chi^2$ (Equation \ref{eq:chi2}) as a function of the scale factor applied to the predicted MP17 profile for propadiene \citep{moses_2017} in fitting the non-auroral mean spectrum at 42$^\circ$N on March 6 2025.  The corresponding 1-mbar abundances are indicated by the upper y-axis and the 1-$\sigma$, 2-$\sigma$ and 3-$\sigma$ confidence levels are indicated by the horizontal and vertical dashed lines.  b) compares the observed spectrum (black points with error bars) with synthetic spectra (solid, colored lines).  The best-fitting spectrum and corresponding propadiene scale factor are indicated in the legend and poorer fitting model spectra are also shown for comparison.  Panels c)-d) and panels e)-f) show similar results for the non-auroral and auroral-mean spectra at 62$^\circ$N, respectively.  }
\label{fig:propadiene_chi2s}
\end{center}
\end{figure*}

\begin{equation}\label{eq:chi2}
    \chi^2 = \sum_i^N \left(\frac{O_i- M_i}{\sigma_i}\right)^2
\end{equation}

$O_i$, $\sigma_i$ and $M_i$ are the observed radiance, uncertainty on observed radiance and modeled radiance at the i$^{th}$ wavenumber, N is the number of wavenumbers.  We considered detection of propadiene statistically significant if $\Delta\chi$ = $(\chi_0^2$ - $\chi_{min}^2)^{0.5} \ge$ 3 (i.e. the 3-$\sigma$ confidence level), where $\chi_0^2$ and $\chi_{min}^2$ are respectively the values calculated using synthetic spectra computed with zero opacity of propadiene and the abundance of propadiene that best fits the observation. Detection was considered tentative if 2 $\le \Delta\chi < $3 (intermediate of the 2-$\sigma$ and 3-$\sigma$ confidence levels) and insignificant if $\Delta\chi <$2.  Figure \ref{fig:propadiene_chi2s} shows the variation in $\chi^2$ with respect to the scale factor applied to the MP17 propadiene profile for three locations and corresponding spectral fits.  

At 42$^\circ$N, an abundance of propadiene that is a factor of 8 higher than the MP17 predicted abundance minimizes $\chi^2$/is the best fitting model (Figure \ref{fig:propadiene_chi2s}a).  However, the synthetic spectrum computed assuming zero opacity of propadiene corresponds to $\Delta \chi \sim$ 2.3: the detection of propadiene at this location is greater than 2$\sigma$ but less than 3$\sigma$, which we consider tentative.  This is further demonstrated in Figure \ref{fig:propadiene_chi2s}b, which compares the observed spectrum at 42$^\circ$N with synthetic spectra: the spectrum computed without propadiene opacity and the spectrum including propadiene at a factor of 8 times higher than predicted both adequately fit the observed spectrum within uncertainty at $\sim$845.25 cm$^{-1}$.  Given the detection is tentative at this location, we quote the abundance at the upper 3-$\sigma$ confidence level as an upper limit, which is a factor of $\sim$19 higher than the MP17 predicted abundance.  This corresponds to a 1-mbar abundance of 1.2 ppbv at the 1-mbar level, where vertical sensitivity to propadiene peaks (Figure \ref{fig:contr_all}).  

However, at higher latitudes, propadiene was detected with $>$3-$\sigma$ statistical significance at all locations. For example, at 62$^\circ$N but sampling longitudes outside the auroral region, the fit to the observed spectra at $\sim$845.25 cm$^{-1}$ was optimized by scaling the MP17 propadiene profile by a factor of 25$^{+5}_{-4}$ (quoting 1-$\sigma$ uncertainties, Figure \ref{fig:propadiene_chi2s}c).   This corresponds to an abundance of 1.6 $\pm$ 0.3 ppbv at the 1-mbar level.  The detection is also spectrally illustrated in Figure \ref{fig:propadiene_chi2s}d, which demonstrates that zero abundance or even the photochemically-predicted abundance of propadiene (a scale factor of 1) produces a featureless spectrum at 845.25 cm$^{-1}$ that does not adequately fit the observed. 

At the same latitude, but sampling longitudes inside the auroral region, propadiene is further enriched: the fit to the observations was optimized using a scale factor of 40 $\pm$ 3 applied to the MP17 propadiene profile (Figure \ref{fig:propadiene_chi2s}e, f), which corresponds to a 1-mbar abundance of 2 $\pm$ 0.2 ppbv.  The synthetic spectrum computed assuming zero opacity of propadiene corresponds to the $\Delta \chi^2 $ = 152 or the 12.3-$\sigma$ level: the detection of propadiene at this latitude in Jupiter's northern auroral region therefore represents a $>$12-$\sigma$ detection.  The fact that propadiene abundances are enriched inside Jupiter's auroral region compared to non-auroral longitudes in the same latitude band and lower latitudes does suggest that the source of the enriched propadiene is the auroral-related chemistry.  We discuss the interpretation of the chemistry further in Section \ref{sec:discuss}.  

\subsection{Detection of propene (C$_3$H$_6$) at $\sim$912.5 cm$^{-1}$}

The 912 cm$^{-1}$ spectra include the emission lines of C$_2$H$_4$. The spectra at 912 cm$^{-1}$ were therefore inverted by adopting the profiles of temperature, NH$_3$, PH$_3$ retrieved in Section \ref{sec:Tp_retr} and allowing the vertical profile of C$_2$H$_4$ to vary.  The predicted MP17 profiles for C$_2$H$_4$ were adopted as \textit{a priori} (Figure \ref{fig:cxhy}, \citealt{moses_2017}) and allowed to vary continuously at all altitudes. 

\begin{figure*}[t!]
\begin{center}
\includegraphics[width=0.99\textwidth]{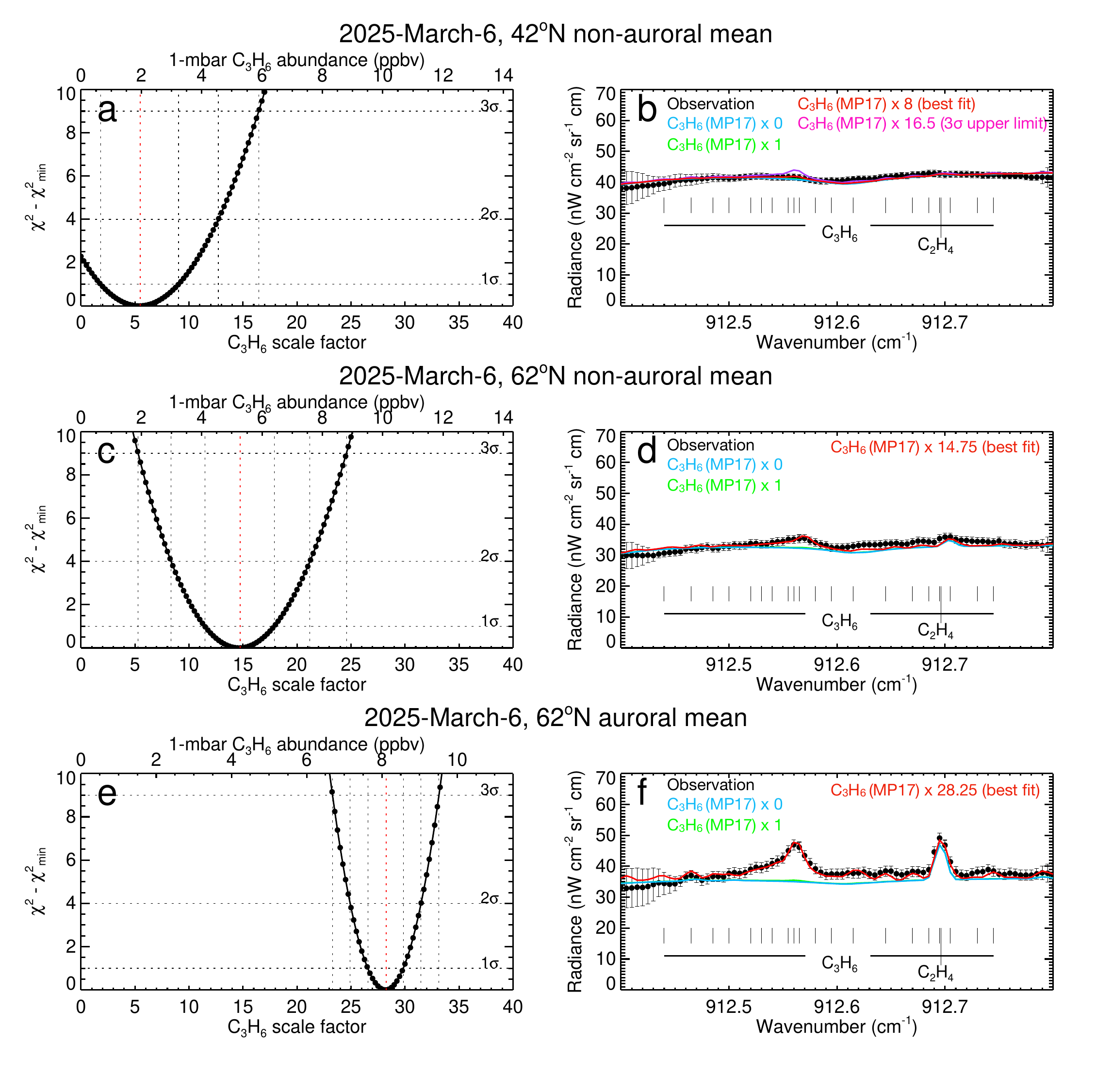}
\caption{a) shows variations in absolute $\chi^2$ (Equation \ref{eq:chi2}) as a function of the scale factor applied to the predicted MP17 profile for propene \citep{moses_2017} in fitting the non-auroral mean spectrum at 42$^\circ$N on 2025-March-6.  The corresponding 1-mbar abundances are indicated by the upper y-axis and the 1-$\sigma$, 2-$\sigma$ and 3-$\sigma$ confidence levels are indicated by the horizontal and vertical dashed lines.  b) compares the observed spectrum (black points with error bars) with synthetic spectra (solid, colored lines).  The best-fitting spectrum and corresponding propene scale factor are indicated in the legend and poorer fitting model spectra are also shown for comparison.  Panels c)-d) and panels e)-f) show similar results for the non-auroral and auroral-mean spectra at 62$^\circ$N, respectively.  The location of the strongest propene and ethylene lines are also indicated.}
\label{fig:propene_chi2s}
\end{center}
\end{figure*}

The retrieved vertical profiles of temperature, NH$_3$, PH$_3$ and C$_2$H$_4$ were adopted and forward model spectra were computed over a grid where the MP17 propene profiles (Figure \ref{fig:cxhy}, \citealt{moses_2017}) were scaled by a constant at all altitudes over a range from 0 to 60.  For each forward model, the $\chi^2$ statistic (Equation \ref{eq:chi2}) was calculated from 912.4 - 912.7 cm$^{-1}$, which covers the expected position of the strongest propene line and surrounding tropospheric continuum.  The detection criteria were the same as those detailed for propadiene in the previous section.  Figure \ref{fig:propene_chi2s} shows the variation in $\chi^2$ with respect to the scale factor applied to the MP17 propene profile for three locations and corresponding spectral fits.

 At 42$^\circ$N, the MP17 propene profile scaled by a factor of 5.5 minimizes the $\chi^2$ value (Figure \ref{fig:propene_chi2s}a).  Zero abundance of propene corresponds to the lower 1.5-$\sigma$ level, which we consider a negligible detection at this location.  This is also demonstrated in Figure \ref{fig:propene_chi2s}b where the forward models assuming scale factors of 0 and 5.5 with respect to the MP17 photochemical profile both fit the spectra with uncertainty at 912.56 cm$^{-1}$.  We therefore quote the abundance at the upper 3-$\sigma$ confidence level as an upper limit, which corresponds to a scale factor of $\sim$16 higher than the MP17 abundance profile or 6 ppbv at the 1-mbar level.  

However, at higher latitudes, and particularly those that overlap with the northern auroral region ($>$55$^\circ$N), the detection of propene is much more significant.  At 62$^\circ$N but sampling longitudes outside the auroral region, the fit to the observed C$_3$H$_6$ emission line was optimized using the predicted MP17 propene profile scaled by a factor of 15$^{+3}_{-4}$ (Figure \ref{fig:propene_chi2s}c).  This corresponds to an abundance of 5.3$^{+1.1}_{-1.2}$ ppbv at the 1-mbar level, where sensitivity to propene peaks (Figure \ref{fig:contr_all}).  This is also demonstrated spectrally in Figure \ref{fig:propene_chi2s}d.  

At 62$^\circ$N, but sampling longitudes inside the auroral region, the abundance of propene is further enriched compared to longitudes outside the auroral region.  The fit was optimized using the MP17 propene profile scaled by a factor of 28 $\pm$ 2 (Figures \ref{fig:propene_chi2s}e), which corresponds to a 1-mbar abundance of 8.1 $\pm$ 0.5 ppbv. The synthetic spectra computed without any propene opacity corresponds to the $\Delta \chi^2$ = 312 or the 17.7-$\sigma$ level.  The detection of propene at this latitude in Jupiter's northern auroral region therefore represents a $>$17-$\sigma$ detection.  The detection is also clearly demonstrated spectrally in Figure \ref{fig:propene_chi2s}f.

\begin{figure*}[t!]
\begin{center}
\includegraphics[width=0.99\textwidth]{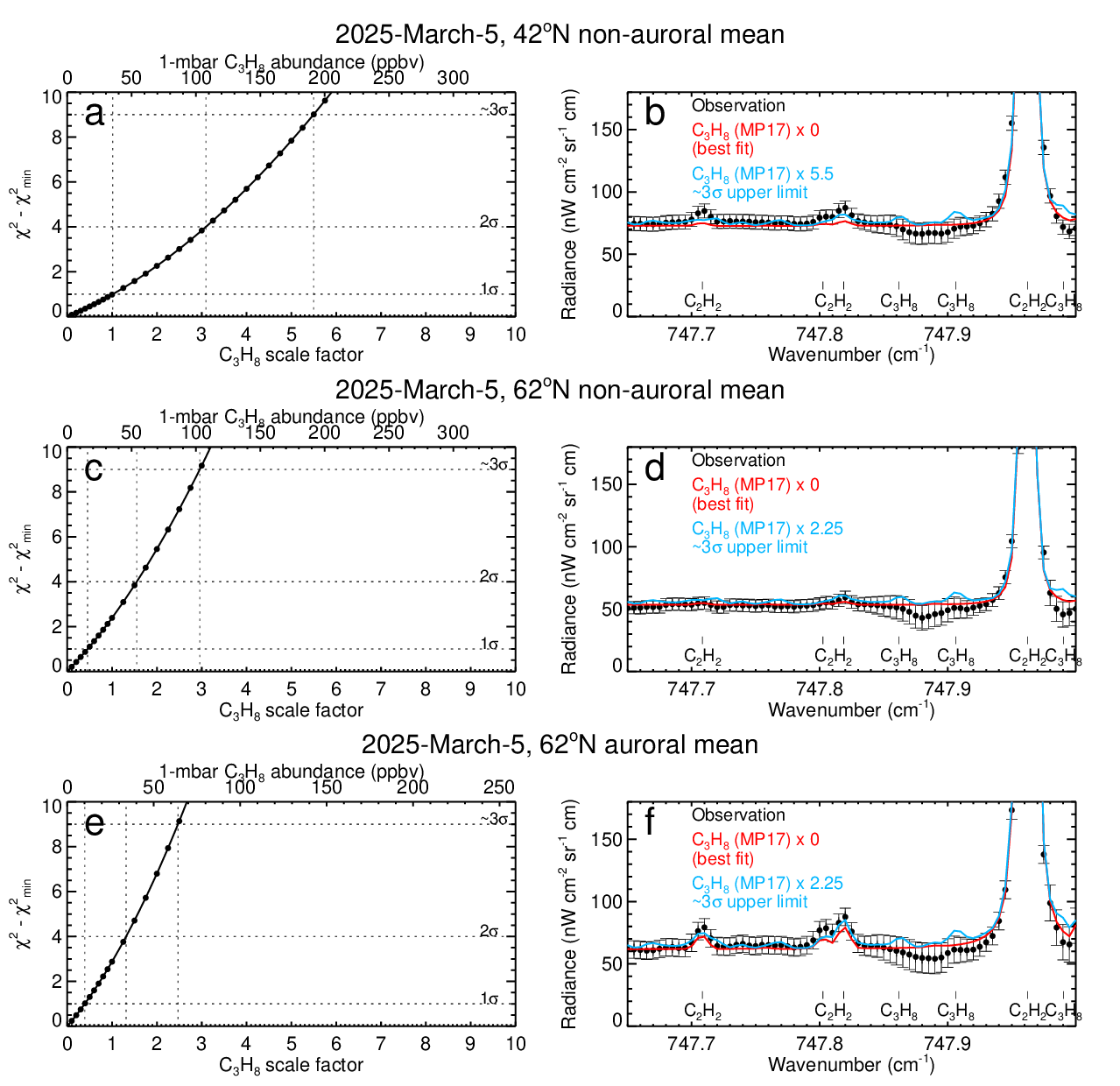}
\caption{a) shows variations in absolute $\chi^2$ (Equation \ref{eq:chi2}) as a function of the scale factor applied to the predicted MP17 profile of propane \citep{moses_2017} in fitting the non-auroral mean spectrum at 42$^\circ$N on 2025-March-6.  The corresponding 10-mbar abundances are indicated by the upper y-axis and the 1-$\sigma$, 2-$\sigma$ and 3-$\sigma$ confidence levels are indicated by the horizontal and vertical dashed lines.  b) compares the observed spectrum (black points with error bars) with synthetic spectra.  The best-fitting spectra (assuming zero abundance of C$_3$H$_8$) and spectra corresponding to the 3-$\sigma$ upper limits are shown indicated in the legend and poorer-fitting model spectra are also shown for comparison.  Panels c)-d) and panels e)-f) show similar results for the non-auroral and auroral-mean spectra at 62$^\circ$N, respectively.}
\label{fig:propane_chi2s}
\end{center}
\end{figure*}

As for propadiene, the fact that propene abundances are enriched inside Jupiter's auroral region compared to non-auroral longitudes in the same latitude band and lower latitudes does suggest that the source of the enrichment is the auroral-related chemistry.  We discuss this further in Section \ref{sec:discuss}.  

\subsection{Non-detection of propane at 748 cm$^{-1}$}

The expected position of propane lines in the 748 cm$^{-1}$ setting neighbor those of the much stronger C$_2$H$_2$ lines.  Adopting the vertical temperature profiles derived in Section \ref{sec:Tp_retr}, the spectra recorded at 748 cm$^{-1}$ were inverted by allowing the vertical profile of C$_2$H$_2$ to vary.  The predicted MP17 profiles of C$_2$H$_2$ (Figure \ref{fig:cxhy}, \citealt{moses_2017}) were respectively adopted as \textit{a priori} and allowed to vary continuously at all altitudes. 

At each location, the retrieved vertical profiles of temperature and C$_2$H$_2$ were adopted, and spectra were forward modelled assuming the predicted MP17 profiles of C$_3$H$_8$ (Figure \ref{fig:cxhy}) scaled at all altitudes by factors ranging from 0 to 100.   Figure \ref{fig:propane_chi2s} shows the $\chi^2$ statistic (Equation \ref{eq:chi2}) calculated for each forward model as a function of C$_3$H$_8$ scale factor for 42$^\circ$N, the non-auroral and auroral mean spectra at 62$^\circ$N. 

At all locations, zero abundance of C$_3$H$_8$ provides the best fit to the observations.  At 42$^\circ$N, we derive a 3-$\sigma$ upper limit abundance of C$_3$H$_8$ that is $\sim$5.5 richer than the MP17 propane profile or approximately 32 ppbv at the 10-mbar level, where sensitivity to stratospheric C$_3$H$_8$ peaks.  At 62$^\circ$N, we derive upper-limit abundances corresponding to $\sim$3 and $\sim$2.5 times photochemically-predicted values outside and inside Jupiter's auroral region, respectively. 

\section{Discussion}\label{sec:discuss}

High-resolution TEXES spectra were recorded in settings centered at 587, 748, 845, 912 and 1248 cm$^{-1}$ on March 5 - 6 2025.  Spectra were sorted into 8-$^\circ$ wide latitude bands.  For latitudes that overlapped with the northern auroral region ($>$55$^\circ$N), spectra were further sorted into those sampling longitudes inside and outside the auroral region (Figure \ref{fig:coadd}.  Spectra in each spatial bin were coadded to improve the signal-to-noise ratio (Figure \ref{fig:example_texes}).  The spectra were then analyzed with the NEMESIS radiative transfer code \citep{irwin_2008} in order to derive atmospheric information and to test for the presence of the targeted propadiene, propene and propane.  We discuss the results derived for temperature and the detectability of propadiene, propene and propane in turn below.

\begin{figure}[t!]
\begin{center}
\includegraphics[width=0.7\textwidth]{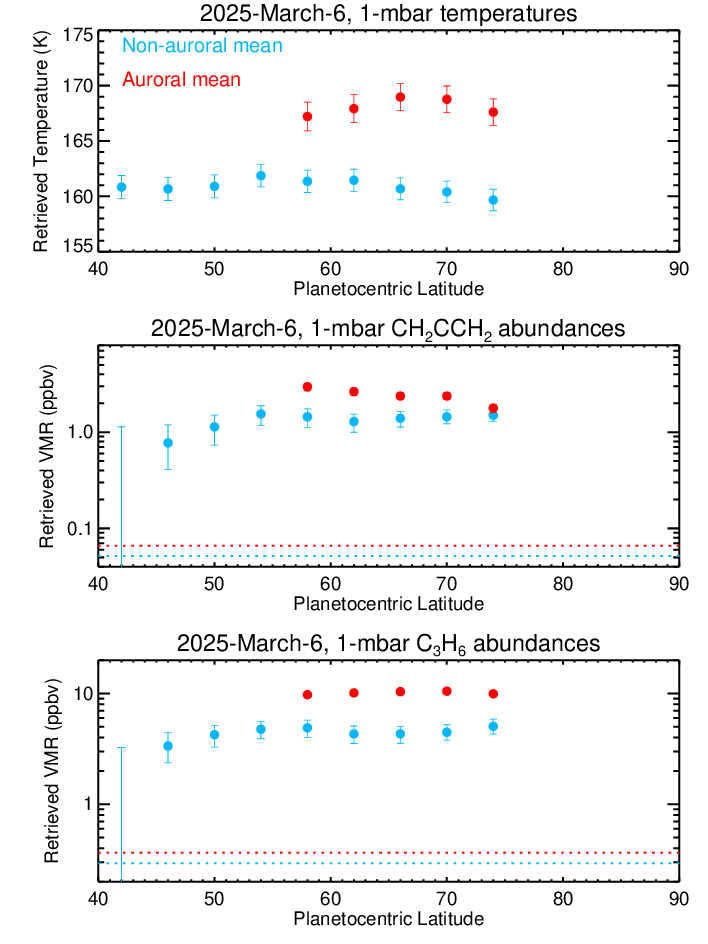}
\caption{Retrieved temperatures (1st panel), and abundances of propadiene (CH$_2$CCH$_2$, 2nd panel) and propene (3rd panel) at the 1-mbar level as a function of planetocentric latitude using observations recorded on March 6 2025.  Note that abundance panels are shown in logarithmic units.  Blue and red results denote temperature/abundances derived from observations that sample outside and inside Jupiter's northern auroral region, respectively.  Error bars are 1 $\sigma$.  Results shown as an upper error bar indicate only a upper limit was derived at that location. Horizontal, dashed lines denote the \textit{a priori} 1-mbar abundances based on neutral photochemical predictions.  }
\label{fig:Tp_cxhy_lat}
\end{center}
\end{figure}

The observations were first inverted to derive the three-dimensional (latitude, longitude, pressure/altitude) temperature distributions, as detailed in Sections \ref{sec:Tp_retr} and \ref{sec:Tp_retr}.  Figure \ref{fig:Tp} shows the temperature distributions retrieved on March 5 and March 6, 2025.  Figure \ref{fig:Tp_cxhy_lat} also shows the retrieved 1-mbar temperatures as a function of latitude.  

As noted in several previous studies (e.g. \citealt{kostiuk_agu_2016,sinclair_2017a,sinclair_2019b,sinclair_2023b,ovalle_2024}), temperatures inside Jupiter’s northern auroral region are predominantly elevated at $\sim$1 mbar, at pressures lower than 0.1 mbar, with relatively cooler temperatures at the intermediate 0.1-mbar level.  As noted in previous work, this pattern in the auroral temperature profile is a robust result that is not an artefact of vertical sensitivity or the chosen a priori profile (e.g. see Figure 8 of \citet{sinclair_2018a} or Figure 6 of \citet{ovalle_2024}). Thermospheric models of Jupiter (e.g. \citealt{bougher_2005,yates_2014}) demonstrate that the upper stratospheric heating (at pressures lower than 0.1 mbar) is expected to be a combination of Joule heating and ion drag heating.  At the time of writing, there are two leading hypotheses for the lower stratospheric auroral-related heating.  First, Juno-MWR (Microwave Radiometer, \citealt{janssen_2005}) observations demonstrate a plasma layer in the lower, auroral stratosphere, which is inferred to result from precipitation of MeV electrons \citep{bhattacharya_2023}.  Second, using ALMA observations, \citet{cavalie_2021} observed a counterotating vortex at $\sim$0.1 mbar coincident with the southern auroral oval.  This was inferred to result from ion-neutral collisions and could represent a lower altitude extension of an electrojet at thermospheric altitudes (e.g. \citealt{achilleos_2001,johnson_2017}).  Time dependent circulation modelling of Jupiter's auroral stratosphere, with parameterization of auroral energy deposition, is required to quantify the potential sources of heating at the 1-mbar level.

The spectra recorded at 748, 845 and 912 cm$^{-1}$ were analyzed to quantify the presence of the spectral features of propane, propadiene and propene, respectively.  The predicted vertical profiles of propadiene, propene or propane \citep{moses_2017} were scaled by a constant at all altitudes and the goodness-of-fit statistic (Equation \ref{eq:chi2}) calculated at the expected wavenumbers of their spectral features.  

The spectral feature of propadiene at 845.25 cm$^{-1}$ was detected with a confidence greater than 12$\sigma$ inside Jupiter's northern auroral region (Figure \ref{fig:propadiene_chi2s}).  At 62$^\circ$N, sampling longitudes outside and inside Jupiter's northern auroral region, we derived abundances that are factors of $\sim$25 and $\sim$40 higher than those predicted by \citealt{moses_2017}, respectively.  This respectively corresponds to abundances of 1.6 $\pm$ 0.3 and 2 $\pm$ 0.2 ppbv at the 1-mbar level, where sensitivity to  propadiene is highest (Figure \ref{fig:contr_all}).  As shown in Figure \ref{fig:Tp_cxhy_lat}, abundances peak at longitudes inside the auroral region, decrease outside the auroral region but in the same latitude circle, and then decrease further towards lower latitudes.  Propadiene is only tentatively detectable in the observations sampling the 42$^\circ$N latitude circle (the lowest latitude sampled by our observations) and so we derived a 3-$\sigma$ upper limit scale factor of $\sim$19 with respect to the photochemical model at that location. This corresponds to a 1-mbar abundance of $\sim$1.2 ppbv.

The spectral features of propene at $\sim$912.56 cm$^{-1}$ were also detected with a confidence greater than 17$\sigma$ in Jupiter's northern auroral region (Figure \ref{fig:propene_chi2s}).  As for propadiene, significant enrichments of propene with respect to neutral photochemical model predictions were required in order to fit the observed spectral features.  At 62$^\circ$N, sampling longitudes outside and inside Jupiter's northern auroral region, we derived abundances that are factors of $\sim$15 and $\sim$28 higher than those predicted by neutral photochemical models \citep{moses_2017}, respectively.  This respectively corresponds to abundances of 5.3$^{+1.1}_{-1.2}$ ppbv and 8.1 $\pm$ 0.5 ppbv at the 1-mbar level, where sensitivity to propene is highest. As for propadiene, derived propene concentrations peak inside the auroral region, decrease outside the auroral region and towards lower latitudes.  Propene was detectable at $>$3$\sigma$ at all locations poleward of 46$^\circ$N whereas the detection at 42$^\circ$N (the lowest latitude sampled by the observations) was negligible (Figure \ref{fig:propene_chi2s}). 

Propane was not detected at any location sampled by the observations (poleward of and including 42$^\circ$N) on March 5 2025.  The absence of spectral features of propane at 747.95 cm$^{-1}$ was used to constrain upper limit abundances.  At 42$^\circ$N, we derive a 3-$\sigma$ upper limit corresponding to the photochemically-predicted abundance of propane \citet{moses_2017} scaled by a factor of 5.5 (Figure \ref{fig:propane_chi2s}).  This corresponds to an abundance of $\sim$19 ppbv at the 10-mbar level, where there is greatest sensitivity to C$_3$H$_8$ (Figure \ref{fig:contr_all}).  At 62$^\circ$N, using a mean of observations sampled outside and inside the northern auroral region, we derived upper limits respectively corresponding to scale factors of $\sim$3 and $\sim$2.5 with respect to the \citet{moses_2017} photochemically-predicted profile.  This corresponds to 10-mbar abundances of $\sim$12 and $\sim$10 ppbv, respectively. Upper limits at higher latitudes are more stringent due to the limb darkening of the tropospheric continuum and limb brightening of stratospheric emissions at higher emission angles. 

The fact that the abundances of propadiene and propene peak inside Jupiter's auroral region does strongly suggest the dominant source of their enrichment is auroral-related chemistry.  The concentrations of other unsaturated hydrocarbons such as acetylene, ethylene, methylacetylene (CH$_3$C$_2$H, an isomer of propadiene), and the aromatic hydrocarbon benzne (C$_6$H$_6$) have been observed to be enriched in Jupiter's auroral region compared to non-auroral locations (e.g. Figure \ref{fig:Tp_cxhy_lat} of this study, \citealt{sinclair_2017a}, \citealt{sinclair_2018a}, \citealt{sinclair_2019b}, \citealt{sinclair_2023b}, \citealt{giles_2023}, \citealt{ovalle_2024}, \citealt{ovalle_c6h6_2024}).  The current interpretation is that auroral-related heating modifies reaction rates and/or exogenous ions and electrons yield ion-neutral and electron recombination reactions at significantly higher rates compared to elsewhere on the planet, which preferentially enrich the abundances of higher-order unsaturated/aromatic hydrocarbons.  Dynamics may also contribute partly or significantly to the observed enhancements.  Downwelling motion would advect air from higher altitudes, which is generally richer in the unsaturated hydrocarbon species (Figure \ref{fig:cxhy}), thereby enriching their abundance at the 1-mbar level where their spectral features peak in sensitivity (Figure \ref{fig:contr_all}).  Furthermore, using ALMA observations, Jupiter's southern auroral stratosphere was demonstrated to be host to a counterrotating vortex at $\sim$0.1 mbar \citep{cavalie_2021}.  If such a circulation also exists in Jupiter's northern auroral stratosphere, chemical constituents could be dynamically confined there thereby augmenting their apparent enrichment.  These results strongly advocate for ion-neutral chemistry and dynamical modelling of Jupiter's auroral atmosphere in order to quantify the relative contributions of the above, proposed mechanisms and to determine the dominant, reaction pathways responsible for enriching the abundances of unsaturated and polycyclic hydrocarbons.  

Surprisingly, the ion chemistry in the auroral region does not enrich the saturated C$_3$ hydrocarbon, propane, with upper limits derived of $\sim$2.5 times those predicted by neutral photochemical models.  Oddly, propane is readily detected in Saturn's atmosphere with abundances that do not vary significantly between auroral and non-auroral regions \citep{greathouse_2006,guerlet_2009,fletcher_poles_2018,fletcher_saturn_2023}.  The lack of detection of propane in the auroral regions here on Jupiter suggests that the unexpectedly high abundances observed at high latitudes on Saturn may result from transport rather than auroral chemistry, making propane a potentially useful tracer of stratospheric circulation on that planet.  Propane is also readily detectable in Titan's atmosphere (e.g. \citealt{maguire_1981,conor_c3h8_2009,lombardo_c3_2019}.  Its undetectability on Jupiter could, in part, be an artefact of vertical sensitivity.  The detectability of propene and propadiene arises both from their aurorally-enriched abundances at Jupiter's high latitudes as well as the vertical sensitivity of their emission features to the $\sim$1-mbar level, where there is strong auroral-related heating warming the line-forming region (Figure \ref{fig:Tp}).  In contrast, the vertical sensitivity to stratospheric propane peaks a decade of pressure higher (Figure \ref{fig:contr_all}), where propane is an order of magnitude less abundant (Figure \ref{fig:cxhy}) and where there is negligible auroral-related heating.  A further explanation could be that the ion-neutral chemical reactions that produce propadiene, propene, and other unsaturated species have rates that are more temperature dependent and/or more sensitive to ion chemistry, compared to those that produce propane.  Furthermore, if vertical motions do contribute to the observed enrichments of propadiene and propene (see discussion in previous paragraph), such vertical motions would be expected to have a negligible or even opposing effect on propane since it lacks a high altitude abundance peak and thus has a negative vertical gradient in abundance in the upper stratosphere (Figure \ref{fig:cxhy}).  However, we note that the vertical profiles in Figure \ref{fig:cxhy} were calculated assuming neutral photochemistry: the vertical profiles of each hydrocarbon could be very different in Jupiter's auroral regions due to the effects of auroral-related heating and chemistry.  Again, the relative roles of chemistry and dynamics in the apparent undetectability of propane in Jupiter's polar stratosphere could be explored with ion-neutral chemistry and dynamical modelling of Jupiter's auroral atmosphere. 

We considered the possibility that the lines at 845.25 cm$^{-1}$ and 912.56 cm$^{-1}$ were actually transitions of another chemical species missing from the spectroscopic line data adopted in our analysis.  However, we ultimately considered this highly unlikely for several reasons.  First, we adopted the most recent line lists of acetylene, ethylene and ethane from HITRAN 2020 \citep{hitran_2020}, which are the dominant emission species of Jupiter's atmosphere in the 800 - 1000 cm$^{-1}$ spectral range.  This minimizes the possibility that the observed lines at 845.25 and 912.56 cm$^{-1}$ are transitions of those species missing from the spectroscopic line data. Second, we reviewed the spectroscopic line data of water (H$_2$O), carbon dioxide (CO$_2$), carbon dioxide (CO), hydroxide (OH), carbonyl sulfide (OCS), formaldehyde (H$_2$CO), deuterated methane (CH$_3$D), hydrogen cyanide (HCN), hydrogen isocyanide (HNC), hydrogen peroxide (H$_2$O$_2$), formic acid (HCOOH), methanol (CH$_3$OH), deuterated acetylene (C$_2$HD), methylcyanide (CH$_3$CN) from HITRAN 2020 \citep{hitran_2020} and/or GEISA 2020 \citep{delahaye_2021} over the spectral ranges of the $\sim$845 and $\sim$912 cm$^{-1}$ settings.  None of these species have significant lines within 0.015 cm$^{-1}$ (the estimated 3-$\sigma$ wavenumber uncertainty corresponding to TEXES' spectral resolving power) of the observed features at 845.25 cm$^{-1}$ and 912.56 cm$^{-1}$.  Furthermore, all these species have stronger, predicted lines located more than 0.015 cm$^{-1}$ from 845.25 cm$^{-1}$ and 912.56 cm$^{-1}$ and no such features are observed at those wavelengths.  We therefore excluded the possibility that the lines at 845.25 cm$^{-1}$ and 912.56 cm$^{-1}$ were one of the aforementioned species.  Third, there is a significant improvement ($\Delta \chi^2 >$ 100 in the correspondence between the synthetic and observed spectra when including the opacities of propadiene at $\sim$845.25 cm$^{-1}$ and propene at 912.56 cm$^{-1}$, which strongly suggests we have correctly identified the species responsible for those lines. 

In future work, we will search for and measure the abundances of propadiene, propene and propane in Jupiter's southern hemisphere.  Given the interpretation that the high abundances of propadiene and propene in Jupiter's northern auroral region are the result of auroral-related chemistry, we would likewise expect similarly high abundances in Jupiter's southern auroral region.  In fact, given the southern auroral region occupies a smaller geographic area and so auroral energy is deposited there with a higher surface density, and given it is approximately co-located with Jupiter's rotational axis that would entrain chemical species there, propadiene and propene may be even more enriched there than in the northern auroral region.  Jupiter's southern auroral region may also prove advantageous for the search for propane given that auroral-related heating there has been demonstrated to reach the deeper ($\sim$10  mbar) pressures \citep{sinclair_2023b} where sensitivity to stratospheric C$_3$H$_8$ is higher (Figure \ref{fig:contr_all}).

\section{Conclusions}\label{sec:conc}

We report a $>$12-$\sigma$ detection of propadiene (CH$_2$CCH$_2$) and a $>$17-$\sigma$ detection of propene (C$_3$H$_6$) in Jupiter's northern auroral stratosphere using IRTF-TEXES spectral measurements.  Spectral scans were performed across Jupiter's mid-to-high northern latitudes on March 5-6 2025 and in settings centered at 587, 748, 843, 912 and 1248 cm$^{-1}$.  The spectra were analyzed with radiative transfer software in order to quantitatively test for the presence of the emission features of propadiene, propene and propane, to derive their abundances if detected, or derive stringent upper limit abundances if not detected.  Both propadiene and propene peak in concentration inside Jupiter's northern auroral region and derived abundances are over an order of magnitude higher than those predicted by neutral photochemical models.  For example, at 62$^\circ$N and sampling longitudes inside Jupiter's northern auroral region, we derived 1-mbar abundances of 2.0 $\pm$ 0.2 ppbv and 8.1 $\pm$ 0.5 ppbv for propadiene and propene, respectively. These abundances are elevated by factors of 40 $\pm$ 3 and 28 $\pm$ 2 with respect to abundances predicted by the \citet{moses_2017} neutral photochemical model.  In contrast, at 42$^\circ$N, both species were only marginally detectable with 3-$\sigma$ upper limits of 1.2 ppbv for propadiene and 6 ppbv for propene at the 1-mbar level.  Spectral features of propane were not detected at any of the locations sampled by the data (poleward of and including 42$^\circ$N).  At 62$^\circ$N and inside Jupiter's northern auroral region, we derive a 3-$\sigma$ C$_3$H$_8$ upper limit of $\sim$10 ppbv at 10 mbar, which is only 2.5 times the \citet{moses_2017} predicted abundance.  The fact that propadiene and propene are most enriched inside Jupiter's auroral region strongly suggests that the heating associated with auroral energy deposition and the influx of ions and electrons modifies the stratospheric chemistry such that unsaturated hydrocarbons are preferentially enriched, as has been observed for the neutral C$_2$ species (e.g. \citealt{sinclair_2017a,sinclair_2018a,sinclair_2019b,sinclair_2023b,giles_2023,ovalle_2024}).  The nondetection of propane could, in part, be explained by the vertical sensitivity of its stratospheric mid-infrared emission lines to deeper $\sim$10 mbar pressures, where it is less abundant and where there is negligible auroral-related heating to warm the line forming region.  The results of this work strongly advocate for development of ion-neutral chemistry models of Jupiter's auroral stratosphere to understand and quantify how magnetospheric particles modify the reaction pathways of higher-order hydrocarbons.

\section{Data Availability Statement}

The IRTF-TEXES data presented in this paper are publicly available at the NASA/IPAC Infared Science Archive: \url{https://irsa.ipac.caltech.edu/applications/irtf/}.  Reduced and calibrated versions of the datasets are available at \url{doi:10.17632/82mncb8fy6.1}.  The photochemical model profiles presented in Figure 4 are available at \url{doi: 10.17632/53z8x7ymyj.1}.  The NEMESIS radiative transfer code, and a python version called nemesisPY, are available publicly at \url{https://github.com/nemesiscode/radtrancode}.   

\section{Acknowledgements}

The research was carried out at the Jet Propulsion Laboratory, California Institute of Technology, under a contract (80NM0018D0004) with the National Aeronautics and Space Administration (NASA).  Data presented in this paper was recorded at the Infrared Telescope Facility, which is operated by the University of Hawaii under contract 80HQTR24DA010 with NASA.  We also want to thank both reviewers for their helpful and constructive feedback.  The High Performance Computing resources used in this investigation were provided by funding from the JPL Enterprise IT Services division.  JM acknowledges support from the NASA grant 80NSSC25K7443.

\newpage
\clearpage
\appendix

\section{Observations}

\begin{table*}[h!]
\footnotesize
\begin{center}
\begin{tabular}{c|c c c c c c c c c c c c|} 
Date & Time & Scan  & Setting     & Slit & R & N$_{spx}$ & CML            &  SOL           &  A & v$_{rad}$ \\
     & (UTC)& number& (cm$^{-1}$) & (``)  &   &          &  ($^\circ$)    &   ($^\circ$)   &    &      (km/s) \\
\hline
& 00:09&8003.01&1248&9.9 x 1.3& 79811 &800&258&2.8&1.82&27.6 \\
& 00:09&8003.02&1248&9.9 x 1.3& 79811 &577&258&2.8&1.82&27.6 \\
& 00:19&8004.01&587&13.3 x 1.9& 46856 &838&276&2.8&1.72&27.7 \\
& 00:19&8004.02&587&13.3 x 1.9& 46856 &911&276&2.8&1.72&27.7 \\
& 00:19&8004.03&587&13.3 x 1.9& 46856 &642&276&2.8&1.72&27.7 \\
& 00:19&8004.04&587&13.3 x 1.9& 46856 &642&276&2.8&1.72&27.7 \\
& 00:30&8005.01&748&12.7 x 1.3& 69149 &670&280&2.8&1.62&27.7 \\
& 00:30&8005.02&748&12.7 x 1.3& 69149 &420&280&2.8&1.62&27.7 \\
& 00:30&8005.03&748&12.7 x 1.3& 69149 &344&280&2.8&1.62&27.7 \\
& 00:30&8005.04&748&12.7 x 1.3& 69149 &459&280&2.8&1.62&27.7 \\
& 00:44&8006.01&748&12.7 x 1.3& 69149 &500&285&2.8&1.51&27.7 \\
& 00:44&8006.02&748&12.7 x 1.3& 69149 &561&285&2.8&1.51&27.7 \\
& 00:44&8006.03&748&12.7 x 1.3& 69149 &584&285&2.8&1.51&27.7 \\
& 00:44&8006.04&748&12.7 x 1.3& 69149 &420&285&2.8&1.51&27.7 \\
& 00:59&8007.01&748&12.7 x 1.3& 69149 &500&291&2.8&1.42&27.7 \\
& 00:59&8007.02&748&12.7 x 1.3& 69149 &529&291&2.8&1.42&27.7 \\
& 00:59&8007.03&748&12.7 x 1.3& 69149 &501&291&2.8&1.42&27.7 \\
& 00:59&8007.04&748&12.7 x 1.3& 69149 &369&291&2.8&1.42&27.7 \\
& 01:24&8010.01&1248&9.9 x 1.3& 79811 &536&314&2.8&1.29&27.7 \\
& 01:24&8010.02&1248&9.9 x 1.3& 79811 &534&314&2.8&1.29&27.7 \\
 & 01:28&8011.01&1248&9.9 x 1.3& 79811 &379&316&2.8&1.27&27.7 \\
& 01:28&8011.02&1248&9.9 x 1.3& 79811 &490&316&2.8&1.27&27.7 \\
& 01:28&8011.03&1248&9.9 x 1.3& 79811 &490&316&2.8&1.27&27.7 \\
& 01:37&8012.01&587&13.3 x 1.9& 46856 &709&319&2.8&1.24&27.7 \\
& 01:37&8012.02&587&13.3 x 1.9& 46856 &938&319&2.8&1.24&27.7 \\
& 01:37&8012.03&587&13.3 x 1.9& 46856 &937&319&2.8&1.24&27.7 \\
& 01:44&8013.01&748&12.7 x 1.3& 69149 &368&322&2.8&1.21&27.8 \\
March 5 2025 & 01:44&8013.02&748&12.7 x 1.3& 69149 &529&322&2.8&1.21&27.8 \\
& 01:44&8013.03&748&12.7 x 1.3& 69149 &570&322&2.8&1.21&27.8 \\
& 01:44&8013.04&748&12.7 x 1.3& 69149 &615&322&2.8&1.21&27.8 \\
& 01:59&8014.01&748&12.7 x 1.3& 69149 &574&327&2.8&1.17&27.8 \\
& 01:59&8014.02&748&12.7 x 1.3& 69149 &684&327&2.8&1.17&27.8 \\
& 01:59&8014.03&748&12.7 x 1.3& 69149 &534&327&2.8&1.17&27.8 \\
& 01:59&8014.04&748&12.7 x 1.3& 69149 &606&327&2.8&1.17&27.8 \\
& 02:14&8015.01&748&12.7 x 1.3& 69149 &445&333&2.8&1.13&27.8 \\
& 02:14&8015.02&748&12.7 x 1.3& 69149 &524&333&2.8&1.13&27.8 \\
& 02:14&8015.03&748&12.7 x 1.3& 69149 &550&333&2.8&1.13&27.8 \\
& 02:23&8016.01&748&12.7 x 1.3& 69149 &648&350&2.8&1.11&27.8 \\
& 02:23&8016.02&748&12.7 x 1.3& 69149 &407&350&2.8&1.11&27.8 \\
& 02:23&8016.03&748&12.7 x 1.3& 69149 &417&350&2.8&1.11&27.8 \\
& 02:23&8016.04&748&12.7 x 1.3& 69149 &503&350&2.8&1.11&27.8 \\
& 02:47&8018.01&1248&9.9 x 1.3& 79811 &665&359&2.8&1.06&27.8 \\
& 02:47&8018.02&1248&9.9 x 1.3& 79811 &507&359&2.8&1.06&27.8 \\
& 02:47&8018.03&1248&9.9 x 1.3& 79811 &507&359&2.8&1.06&27.8 \\
& 02:47&8018.04&1248&9.9 x 1.3& 79811 &507&359&2.8&1.06&27.8 \\
& 02:59&8019.01&587&13.3 x 1.9& 46856 &1072&3&2.8&1.04&27.9 \\
& 02:59&8019.02&587&13.3 x 1.9& 46856 &584&3&2.8&1.04&27.9 \\
& 02:59&8019.03&587&13.3 x 1.9& 46856 &584&3&2.8&1.04&27.9 \\
& 03:06&8020.01&748&12.7 x 1.3& 69149 &531&6&2.8&1.03&27.9 \\
& 03:06&8020.02&748&12.7 x 1.3& 69149 &257&6&2.8&1.03&27.9 \\
& 03:06&8020.03&748&12.7 x 1.3& 69149 &251&6&2.8&1.03&27.9 \\
& 03:41&8021.01&748&12.7 x 1.3& 69149 &564&33&2.8&1.00&27.9 \\
& 03:41&8021.02&748&12.7 x 1.3& 69149 &556&33&2.8&1.00&27.9 \\
& 03:41&8021.03&748&12.7 x 1.3& 69149 &670&33&2.8&1.00&27.9 \\
& 03:41&8021.04&748&12.7 x 1.3& 69149 &615&33&2.8&1.00&27.9 \\
 \hline
\end{tabular}
\end{center}
\caption{Details of the IRTF-TEXES observations presented in this work.  Dates/times are UTC, the slit dimensions (length x width) are in arcseconds, R is the resolving power ($\nu/\Delta\nu$), the central meridian longitude (CML) at the time of scan is System III and the sub-solar latitude (SOL) is planetocentric. A is the mean airmass throughout the scan and v$_{rad}$ is the relative Earth-Jupiter velocity in km/s. }
\label{tab:texes_obs1}
\end{table*}

\setcounter{table}{0}
\begin{table*}[h!]
\footnotesize
\begin{center}
\begin{tabular}{c|c c c c c c c c c c c c|} 
Date & Time & Scan  & Setting     & Slit & R & N$_{spx}$ & CML            &  SOL           &  A & v$_{rad}$ \\
     & (UTC)& number& (cm$^{-1}$) & (``)  &   &          &  ($^\circ$)    &   ($^\circ$)   &    &      (km/s) \\
\hline
& 03:56&8022.01&748&12.7 x 1.3& 69149 &707&39&2.8&1.00&27.9 \\
& 03:56&8022.02&748&12.7 x 1.3& 69149 &583&39&2.8&1.00&27.9 \\
& 03:56&8022.03&748&12.7 x 1.3& 69149 &701&39&2.8&1.00&27.9 \\
& 03:56&8022.04&748&12.7 x 1.3& 69149 &583&39&2.8&1.00&27.9 \\
& 04:11&8023.01&748&12.7 x 1.3& 69149 &719&44&2.8&1.00&28.0 \\
& 04:11&8023.02&748&12.7 x 1.3& 69149 &460&44&2.8&1.00&28.0 \\
& 04:11&8023.03&748&12.7 x 1.3& 69149 &583&44&2.8&1.00&28.0 \\
& 04:11&8023.04&748&12.7 x 1.3& 69149 &577&44&2.8&1.00&28.0 \\
& 04:31&8025.01&1248&9.9 x 1.3& 79811 &236&66&2.8&1.00&28.0 \\
& 04:31&8025.02&1248&9.9 x 1.3& 79811 &529&66&2.8&1.00&28.0 \\
& 04:31&8025.03&1248&9.9 x 1.3& 79811 &533&66&2.8&1.00&28.0 \\
& 04:40&8026.01&587&13.3 x 1.9& 46856 &746&69&2.8&1.00&28.0 \\
& 04:40&8026.02&587&13.3 x 1.9& 46856 &746&69&2.8&1.00&28.0 \\
& 04:40&8026.03&587&13.3 x 1.9& 46856 &761&69&2.8&1.00&28.0 \\
& 04:47&8027.01&748&12.7 x 1.3& 69149 &386&72&2.8&1.01&28.0 \\
& 04:47&8027.02&748&12.7 x 1.3& 69149 &465&72&2.8&1.01&28.0 \\
& 04:47&8027.03&748&12.7 x 1.3& 69149 &515&72&2.8&1.01&28.0 \\
& 04:47&8027.04&748&12.7 x 1.3& 69149 &622&72&2.8&1.01&28.0 \\
& 05:01&8028.01&748&12.7 x 1.3& 69149 &659&77&2.8&1.02&28.1 \\
& 05:01&8028.02&748&12.7 x 1.3& 69149 &699&77&2.8&1.02&28.1 \\
& 05:01&8028.03&748&12.7 x 1.3& 69149 &617&77&2.8&1.02&28.1 \\
& 05:01&8028.04&748&12.7 x 1.3& 69149 &537&77&2.8&1.02&28.1 \\
& 05:15&8029.01&748&12.7 x 1.3& 69149 &594&96&2.8&1.03&28.1 \\
& 05:15&8029.02&748&12.7 x 1.3& 69149 &593&96&2.8&1.03&28.1 \\
& 05:19&8030.01&748&12.7 x 1.3& 69149 &617&98&2.8&1.03&28.1 \\
& 05:19&8030.02&748&12.7 x 1.3& 69149 &617&98&2.8&1.03&28.1 \\
& 05:19&8030.03&748&12.7 x 1.3& 69149 &675&98&2.8&1.03&28.1 \\
March 5 2025 & 05:19&8030.04&748&12.7 x 1.3& 69149 &537&98&2.8&1.03&28.1 \\
& 05:41&8032.01&1248&9.9 x 1.3& 79811 &570&106&2.8&1.06&28.1 \\
& 05:41&8032.02&1248&9.9 x 1.3& 79811 &302&106&2.8&1.06&28.1 \\
& 05:41&8032.03&1248&9.9 x 1.3& 79811 &302&106&2.8&1.06&28.1 \\
& 05:49&8033.01&587&13.3 x 1.9& 46856 &581&109&2.8&1.08&28.2 \\
& 05:49&8033.02&587&13.3 x 1.9& 46856 &785&109&2.8&1.08&28.2 \\
& 05:49&8033.03&587&13.3 x 1.9& 46856 &814&109&2.8&1.08&28.2 \\
& 05:56&8034.01&748&12.7 x 1.3& 69149 &498&111&2.8&1.09&28.2 \\
& 05:56&8034.02&748&12.7 x 1.3& 69149 &498&111&2.8&1.09&28.2 \\
& 06:01&8035.01&748&12.7 x 1.3& 69149 &461&127&2.8&1.10&28.2 \\
& 06:01&8035.02&748&12.7 x 1.3& 69149 &461&127&2.8&1.10&28.2 \\
& 06:27&8036.01&748&12.7 x 1.3& 69149 &737&137&2.8&1.17&28.2 \\
& 06:27&8036.02&748&12.7 x 1.3& 69149 &568&137&2.8&1.17&28.2 \\
& 06:27&8036.03&748&12.7 x 1.3& 69149 &592&137&2.8&1.17&28.2 \\
& 06:58&8037.01&748&12.7 x 1.3& 69149 &551&148&2.8&1.27&28.3 \\
& 06:58&8037.02&748&12.7 x 1.3& 69149 &592&148&2.8&1.27&28.3 \\
& 06:58&8037.03&748&12.7 x 1.3& 69149 &632&148&2.8&1.27&28.3 \\
& 06:58&8037.04&748&12.7 x 1.3& 69149 &719&148&2.8&1.27&28.3 \\
& 07:24&8040.01&1248&9.9 x 1.3& 79811 &618&172&2.8&1.39&28.3 \\
& 07:24&8040.02&1248&9.9 x 1.3& 79811 &618&172&2.8&1.39&28.3 \\
& 07:28&8041.01&1248&9.9 x 1.3& 79811 &336&174&2.8&1.42&28.3 \\
& 07:28&8041.02&1248&9.9 x 1.3& 79811 &373&174&2.8&1.42&28.3 \\
& 07:33&8042.01&587&13.3 x 1.9& 46856 &711&175&2.8&1.45&28.3 \\
& 07:33&8042.02&587&13.3 x 1.9& 46856 &734&175&2.8&1.45&28.3 \\
& 07:33&8042.03&587&13.3 x 1.9& 46856 &723&175&2.8&1.45&28.3 \\
 & 07:40&8043.01&748&12.7 x 1.3& 69149 &473&178&2.8&1.50&28.3 \\
& 07:40&8043.02&748&12.7 x 1.3& 69149 &434&178&2.8&1.50&28.3 \\
& 07:46&8044.01&748&12.7 x 1.3& 69149 &528&180&2.8&1.54&28.3 \\
& 07:46&8044.02&748&12.7 x 1.3& 69149 &528&180&2.8&1.54&28.3 \\
& 07:46&8044.03&748&12.7 x 1.3& 69149 &551&180&2.8&1.54&28.3 \\
 \hline
\end{tabular}
\end{center}
\caption{(continued from previous page)}
\end{table*}

\setcounter{table}{0}
\begin{table*}[h!]
\footnotesize
\begin{center}
\begin{tabular}{c|c c c c c c c c c c c c|} 
Date & Time & Scan  & Setting     & Slit & R & N$_{spx}$ & CML            &  SOL           &  A & v$_{rad}$ \\
     & (UTC)& number& (cm$^{-1}$) & (``)  &   &          &  ($^\circ$)    &   ($^\circ$)   &    &      (km/s) \\
\hline
& 07:46&8044.04&748&12.7 x 1.3& 69149 &512&180&2.8&1.54&28.3 \\
& 08:01&8045.01&748&12.7 x 1.3& 69149 &551&200&2.8&1.66&28.4 \\
& 08:01&8045.02&748&12.7 x 1.3& 69149 &473&200&2.8&1.66&28.4 \\
& 08:01&8045.03&748&12.7 x 1.3& 69149 &561&200&2.8&1.66&28.4 \\
& 08:01&8045.04&748&12.7 x 1.3& 69149 &466&200&2.8&1.66&28.4 \\
& 08:17&8046.01&748&12.7 x 1.3& 69149 &621&206&2.8&1.81&28.4 \\
& 08:17&8046.02&748&12.7 x 1.3& 69149 &544&206&2.8&1.81&28.4 \\
March 5 2025 & 08:17&8046.03&748&12.7 x 1.3& 69149 &599&206&2.8&1.81&28.4 \\
& 08:17&8046.04&748&12.7 x 1.3& 69149 &602&206&2.8&1.81&28.4 \\
& 08:29&8047.01&1248&9.9 x 1.3& 79811 &268&210&2.8&1.96&28.4 \\
& 08:29&8047.02&1248&9.9 x 1.3& 79811 &420&210&2.8&1.96&28.4 \\
& 08:29&8047.03&1248&9.9 x 1.3& 79811 &404&210&2.8&1.96&28.4 \\
& 08:29&8047.04&1248&9.9 x 1.3& 79811 &404&210&2.8&1.96&28.4 \\
& 08:41&8048.01&587&13.3 x 1.9& 46856 &759&215&2.8&2.13&28.4 \\
& 08:41&8048.02&587&13.3 x 1.9& 46856 &789&215&2.8&2.13&28.4 \\
& 08:41&8048.03&587&13.3 x 1.9& 46856 &801&215&2.8&2.13&28.4 \\
\hline
& 00:07&9002.01&1248&9.9 x 1.3& 79811 &700&48&2.8&1.81&27.6 \\
& 00:07&9002.02&1248&9.9 x 1.3& 79811 &381&48&2.8&1.81&27.6 \\
& 00:07&9002.03&1248&9.9 x 1.3& 79811 &420&48&2.8&1.81&27.6 \\
& 00:18&9003.01&587&13.3 x 1.9& 46856 &403&66&2.8&1.69&27.6 \\
& 00:18&9003.02&587&13.3 x 1.9& 46856 &604&66&2.8&1.69&27.6 \\
& 00:18&9003.03&587&13.3 x 1.9& 46856 &604&66&2.8&1.69&27.6 \\
& 00:25&9004.01&843&8.50 x 1.3&72406&460&69&2.8&1.63&27.7 \\
& 00:25&9004.02&843&8.50 x 1.3&72406&381&69&2.8&1.63&27.7 \\
& 00:25&9004.03&843&8.50 x 1.3&72406&460&69&2.8&1.63&27.7 \\
& 00:25&9004.04&843&8.50 x 1.3&72406&445&69&2.8&1.63&27.7 \\
& 00:36&9005.01&912&11.7 x 1.3& 74229 &308&73&2.8&1.54&27.7 \\
& 00:36&9005.02&912&11.7 x 1.3& 74229 &385&73&2.8&1.54&27.7 \\
& 00:36&9005.03&912&11.7 x 1.3& 74229 &462&73&2.8&1.54&27.7 \\
& 00:36&9005.04&912&11.7 x 1.3& 74229 &501&73&2.8&1.54&27.7 \\
& 00:59&9007.01&1248&9.9 x 1.3& 79811 &370&81&2.8&1.40&27.7 \\
& 00:59&9007.02&1248&9.9 x 1.3& 79811 &367&81&2.8&1.40&27.7 \\
& 01:09&9008.01&587&13.3 x 1.9& 46856 &346&85&2.8&1.35&27.7 \\
& 01:09&9008.02&587&13.3 x 1.9& 46856 &818&85&2.8&1.35&27.7 \\
& 01:16&9009.01&843&8.50 x 1.3&72406&402&102&2.8&1.31&27.7 \\
& 01:16&9009.02&843&8.50 x 1.3&72406&402&102&2.8&1.31&27.7 \\
& 01:16&9009.03&843&8.50 x 1.3&72406&360&102&2.8&1.31&27.7 \\
& 01:16&9009.04&843&8.50 x 1.3&72406&330&102&2.8&1.31&27.7 \\
March 6 2025 & 01:26&9010.01&843&8.50 x 1.3&72406&402&106&2.8&1.27&27.7 \\
 & 01:26&9010.02&843&8.50 x 1.3&72406&330&106&2.8&1.27&27.7 \\
& 01:26&9010.03&843&8.50 x 1.3&72406&402&106&2.8&1.27&27.7 \\
& 01:26&9010.04&843&8.50 x 1.3&72406&383&106&2.8&1.27&27.7 \\
& 01:38&9011.01&912&11.7 x 1.3& 74229 &542&110&2.8&1.22&27.7 \\
& 01:38&9011.02&912&11.7 x 1.3& 74229 &352&110&2.8&1.22&27.7 \\
& 01:38&9011.03&912&11.7 x 1.3& 74229 &423&110&2.8&1.22&27.7 \\
& 01:38&9011.04&912&11.7 x 1.3& 74229 &423&110&2.8&1.22&27.7 \\
& 01:48&9012.01&912&11.7 x 1.3& 74229 &390&114&2.8&1.19&27.7 \\
& 01:48&9012.02&912&11.7 x 1.3& 74229 &412&114&2.8&1.19&27.7 \\
& 01:48&9012.03&912&11.7 x 1.3& 74229 &491&114&2.8&1.19&27.7 \\
& 01:48&9012.04&912&11.7 x 1.3& 74229 &494&114&2.8&1.19&27.7 \\
& 02:13&9015.01&1248&9.9 x 1.3& 79811 &523&137&2.8&1.12&27.8 \\
& 02:13&9015.02&1248&9.9 x 1.3& 79811 &384&137&2.8&1.12&27.8 \\
& 02:22&9016.01&587&13.3 x 1.9& 46856 &714&140&2.8&1.10&27.8 \\
& 02:22&9016.02&587&13.3 x 1.9& 46856 &782&140&2.8&1.10&27.8 \\
& 02:28&9017.01&843&8.50 x 1.3&72406&256&143&2.8&1.09&27.8 \\
& 02:28&9017.02&843&8.50 x 1.3&72406&402&143&2.8&1.09&27.8 \\
& 02:28&9017.03&843&8.50 x 1.3&72406&402&143&2.8&1.09&27.8 \\
& 02:28&9017.04&843&8.50 x 1.3&72406&383&143&2.8&1.09&27.8 \\
& 02:39&9018.01&843&8.50 x 1.3&72406&459&147&2.8&1.07&27.8 \\
& 02:39&9018.02&843&8.50 x 1.3&72406&365&147&2.8&1.07&27.8 \\
& 02:39&9018.03&843&8.50 x 1.3&72406&383&147&2.8&1.07&27.8 \\
& 02:39&9018.04&843&8.50 x 1.3&72406&342&147&2.8&1.07&27.8 \\
 \hline
\end{tabular}
\end{center}
\caption{(continued from previous page)}
\end{table*}

\setcounter{table}{1}
\begin{table*}[h!]
\footnotesize
\begin{center}
\begin{tabular}{c|c c c c c c c c c c c c|} 
Date & Time & Scan  & Setting     & Slit & R & N$_{spx}$ & CML            &  SOL           &  A & v$_{rad}$ \\
     & (UTC)& number& (cm$^{-1}$) & (``)  &   &          &  ($^\circ$)    &   ($^\circ$)   &    &      (km/s) \\
\hline
& 02:50&9019.01&912&11.7 x 1.3& 74229 &710&151&2.8&1.05&27.8 \\
& 02:50&9019.02&912&11.7 x 1.3& 74229 &611&151&2.8&1.05&27.8 \\
& 02:50&9019.03&912&11.7 x 1.3& 74229 &634&151&2.8&1.05&27.8 \\
& 02:50&9019.04&912&11.7 x 1.3& 74229 &533&151&2.8&1.05&27.8 \\
& 03:01&9020.01&912&11.7 x 1.3& 74229 &491&155&2.8&1.04&27.8 \\
& 03:01&9020.02&912&11.7 x 1.3& 74229 &531&155&2.8&1.04&27.8 \\
& 03:01&9020.03&912&11.7 x 1.3& 74229 &496&155&2.8&1.04&27.8 \\
& 03:01&9020.04&912&11.7 x 1.3& 74229 &531&155&2.8&1.04&27.8 \\
& 03:12&9021.01&1248&9.9 x 1.3& 79811 &267&159&2.8&1.02&27.8 \\
& 03:12&9021.02&1248&9.9 x 1.3& 79811 &335&159&2.8&1.02&27.8 \\
& 03:21&9022.01&587&13.3 x 1.9& 46856 &829&176&2.8&1.02&27.9 \\
& 03:21&9022.02&587&13.3 x 1.9& 46856 &850&176&2.8&1.02&27.9 \\
& 03:27&9023.01&843&8.50 x 1.3&72406&364&178&2.8&1.01&27.9 \\
& 03:27&9023.02&843&8.50 x 1.3&72406&329&178&2.8&1.01&27.9 \\
& 03:27&9023.03&843&8.50 x 1.3&72406&294&178&2.8&1.01&27.9 \\
& 03:27&9023.04&843&8.50 x 1.3&72406&458&178&2.8&1.01&27.9 \\
& 03:38&9024.01&843&8.50 x 1.3&72406&382&182&2.8&1.00&27.9 \\
& 03:38&9024.02&843&8.50 x 1.3&72406&438&182&2.8&1.00&27.9 \\
& 03:38&9024.03&843&8.50 x 1.3&72406&329&182&2.8&1.00&27.9 \\
& 03:38&9024.04&843&8.50 x 1.3&72406&329&182&2.8&1.00&27.9 \\
& 03:50&9025.01&912&11.7 x 1.3& 74229 &781&187&2.8&1.00&27.9 \\
& 03:50&9025.02&912&11.7 x 1.3& 74229 &607&187&2.8&1.00&27.9 \\
& 03:50&9025.03&912&11.7 x 1.3& 74229 &626&187&2.8&1.00&27.9 \\
& 03:50&9025.04&912&11.7 x 1.3& 74229 &661&187&2.8&1.00&27.9 \\
& 04:01&9026.01&912&11.7 x 1.3& 74229 &597&191&2.8&1.00&27.9 \\
& 04:01&9026.02&912&11.7 x 1.3& 74229 &655&191&2.8&1.00&27.9 \\
& 04:01&9026.03&912&11.7 x 1.3& 74229 &629&191&2.8&1.00&27.9 \\
March 6 2025  & 04:01&9026.04&912&11.7 x 1.3& 74229 &557&191&2.8&1.00&27.9 \\
& 04:20&9028.01&1248&9.9 x 1.3& 79811 &550&212&2.8&1.00&28.0 \\
& 04:20&9028.02&1248&9.9 x 1.3& 79811 &556&212&2.8&1.00&28.0 \\
& 04:24&9029.01&1248&9.9 x 1.3& 79811 &472&214&2.8&1.00&28.0 \\
& 04:24&9029.02&1248&9.9 x 1.3& 79811 &434&214&2.8&1.00&28.0 \\
& 04:33&9030.01&587&13.3 x 1.9& 46856 &835&217&2.8&1.00&28.0 \\
& 04:33&9030.02&587&13.3 x 1.9& 46856 &811&217&2.8&1.00&28.0 \\
& 04:40&9031.01&843&8.50 x 1.3&72406&260&219&2.8&1.00&28.0 \\
& 04:40&9031.02&843&8.50 x 1.3&72406&363&219&2.8&1.00&28.0 \\
& 04:40&9031.03&843&8.50 x 1.3&72406&419&219&2.8&1.00&28.0 \\
& 04:40&9031.04&843&8.50 x 1.3&72406&328&219&2.8&1.00&28.0 \\
& 04:50&9032.01&843&8.50 x 1.3&72406&400&223&2.8&1.01&28.0 \\
& 04:50&9032.02&843&8.50 x 1.3&72406&419&223&2.8&1.01&28.0 \\
& 04:50&9032.03&843&8.50 x 1.3&72406&490&223&2.8&1.01&28.0 \\
& 04:50&9032.04&843&8.50 x 1.3&72406&363&223&2.8&1.01&28.0 \\
& 05:02&9033.01&912&11.7 x 1.3& 74229 &598&227&2.8&1.02&28.0 \\
& 05:02&9033.02&912&11.7 x 1.3& 74229 &557&227&2.8&1.02&28.0 \\
& 05:02&9033.03&912&11.7 x 1.3& 74229 &557&227&2.8&1.02&28.0 \\
& 05:02&9033.04&912&11.7 x 1.3& 74229 &648&227&2.8&1.02&28.0 \\
& 05:13&9034.01&912&11.7 x 1.3& 74229 &479&231&2.8&1.03&28.1 \\
& 05:13&9034.02&912&11.7 x 1.3& 74229 &579&231&2.8&1.03&28.1 \\
& 05:13&9034.03&912&11.7 x 1.3& 74229 &577&231&2.8&1.03&28.1 \\
& 05:13&9034.04&912&11.7 x 1.3& 74229 &539&231&2.8&1.03&28.1 \\
& 05:25&9035.01&1248&9.9 x 1.3& 79811 &326&250&2.8&1.04&28.1 \\
& 05:25&9035.02&1248&9.9 x 1.3& 79811 &550&250&2.8&1.04&28.1 \\
& 05:25&9035.03&1248&9.9 x 1.3& 79811 &529&250&2.8&1.04&28.1 \\
& 05:34&9036.01&587&13.3 x 1.9& 46856 &1036&254&2.8&1.06&28.1 \\
& 05:34&9036.02&587&13.3 x 1.9& 46856 &932&254&2.8&1.06&28.1 \\
& 05:34&9036.03&587&13.3 x 1.9& 46856 &898&254&2.8&1.06&28.1 \\
 \hline
\end{tabular}
\end{center}
\caption{(continued from previous page)}
\end{table*}

\end{document}